\documentclass[
reprint,
bibnotes,
 amsmath,amssymb,
 aps,
]{revtex4-2}

\usepackage{graphicx}
\usepackage{dcolumn}
\usepackage{bm}
\usepackage{float}
\usepackage[many]{tcolorbox}
\usepackage{shuffle}
\usepackage{graphbox}
\usepackage{environ}
\usepackage{physics}
\usepackage{tabularx}
\usepackage{diagbox}
\usepackage{subcaption}
\usepackage{multirow}

\begin{document}

\title{Covariant color-kinematics duality, Hopf algebras and permutohedra}
%

\author{Qu Cao$^{1,5}$}
\email{qucao@zju.edu.cn}
\author{Jin Dong$^{1,4}$}
\email{dongjin@itp.ac.cn}
\author{Song He$^{1,2,3}$}
\email{songhe@itp.ac.cn}
\author{Yao-Qi Zhang$^{1,4}$}
\email{zhangyaoqi@itp.ac.cn}
\affiliation{$^{1}$CAS Key Laboratory of Theoretical Physics, Institute of Theoretical Physics, Chinese Academy of Sciences, Beijing 100190, China \\
$^{2}$School of Fundamental Physics and Mathematical Sciences, Hangzhou Institute for Advanced Study;\\
International Centre for Theoretical Physics Asia-Pacific, Beijing/Hangzhou, China\\
$^{3}$Peng Huanwu Center for Fundamental Theory, Hefei, Anhui 230026, P. R. China\\
$^{4}$School of Physical Sciences, University of Chinese Academy of Sciences, No.19A Yuquan Road, Beijing 100049, China\\
$^{5}$Zhejiang Institute of Modern Physics, Department of Physics, Zhejiang University, Hangzhou, 310027, China
}\date{\today}

\begin{abstract}
Based on the covariant color-kinematics duality, we investigate combinatorial and algebraic structures underlying their Bern-Carrasco-Johansson (BCJ) numerators of tree-level amplitudes in Yang-Mills-scalar (YMS) theory. The closed-formulae for BCJ numerators of YMS amplitudes and the pure-Yang-Mills (YM) ones exhibit nice quasi-shuffle Hopf algebra structures, and interestingly they can be viewed as summing over boundaries of all dimensions of a combinatorial {\it permutohedron}. In particular, the numerator with two scalars and $n{-}2$ gluons contains Fubini number ($\mathcal{F}_{n{-}2}$) of terms in one-to-one correspondence with boundaries of a $(n{-}3)$-dimensional permutohedron, and each of them has its own spurious-pole structures and a gauge-invariant numerator (both depending on reference momenta). From such Hopf algebra or permutohedron structure, we derive new recursion relations for the numerators and intriguing ``factorization” on each spurious pole/facet of the permutohedron. Similar results hold for general YMS numerators and the pure-YM ones. Finally, with a special choice of reference momenta, our results imply BCJ numerators in a heavy-mass effective field theory with two massive particles and $n{-}2$ gluons/gravitons: we observe highly nontrivial cancellations in the heavy-mass limit, leading to new formulae for the effective numerators which resemble those obtained in recent works. 
\end{abstract}
\maketitle


\section{Introduction}
Despite very different natures, gauge theories and gravity have deep connections; one of the oldest and the most prominent example is the double copy structure \cite{Kawai:1985xq, Bern:2008qj, Bern:2010ue}. Originally it was discovered from Kawai-Lewellen-Tye (KLT) relations \cite{Kawai:1985xq} in string theory, and a modern realization of double copy has relied on the duality between color and kinematics for gauge theory amplitudes, where the Bern-Carrasco-Johansson (BCJ) kinematic numerators satisfy the same Jacobi relations as the color factors. The duality and double copy have led to tremendous progress in the study of amplitudes both in gauge theory and gravity (see~\cite{Bern:2019prr, Bern:2022wqg, Adamo:2022dcm} and references therein). More recently, the authors of~\cite{Cheung:2021zvb} have revealed a so-called covariant color-kinematics (CCK) duality for a large class of theories including Yang-Mills theory (YM) and its coupling to bi-adjoint $\phi^3$ (YMS). As a consequence, the duality implies new, closed-form expression for BCJ numerators of all tree-level amplitudes in YMS and YM theory. 
Previous works on BCJ numerators and kinematic algebras include~\cite{Mafra:2011kj,Bargheer:2012gv,CHY3,He:2015wgf,Fu:2017uzt,Teng:2017tbo, Du:2017kpo, He:2017spx, Edison:2020ehu,He:2021lro, Monteiro:2011pc,Monteiro:2013rya,Cheung:2016prv,Chen:2019ywi,Edison:2022jln} and references therein. 

On the other hand, recent years have seen progress on revealing new geometric/combinatorial structures underlying scattering amplitudes {\it e.g.} from the (all-loop) amplituhedron of supersymmetric Yang-Mills ~\cite{Arkani-Hamed:2013jha} to the associahedron for bi-adjoint $\phi^3$ at tree level~\cite{Arkani-Hamed:2017mur} (with extensions to string scattering~\cite{Arkani-Hamed:2019mrd}). It is natural to look for hints of such structures underlying YM and gravity amplitudes; instead of directly working with tree amplitudes, one may decompose the problem and ask a somewhat strange question as a first step: are there combinatorial structures underlying BCJ numerators?

In this note, we take BCJ numerators from CCK duality~\cite{Cheung:2021zvb} as inputs and present preliminary evidence for such structures: in addition to the more familiar quasi-shuffle Hopf algebras~\cite{Hoffman}, we find hidden combinatorial permutohedra \cite{wiki:Permutohedron} for BCJ numerators. Any BCJ numerator can be written as the sum over all boundaries of a permutohedron (or terms from a quasi-shuffle product); for a co-dimension $d$ boundary (length-$d$ term), it contains a product of $d{+}1$ factors each with a spurious pole and a gauge-invariant numerator. We will focus on the case with two scalars and $n{-}2$ gluons, which corresponds to a $(n{-}3)$-dimensional permutohedron, and it has Fubini number ${\cal F}_{n{-}2}$ boundaries with co-dimensions $d=0, 1, \ldots, n{-}3$; each boundary is labeled by $d+1$ subsets, and for each factor labeled by such a set both the numerator (which is gauge invariant in the gluons) and the (spurious pole) denominator are given by Lorentz products of momenta and polarizations, as well as the reference momenta. Apart from being the most illustrative BCJ numerators of YMS cases, we will also see that they give nice BCJ numerators in the heavy-mass effective theory (HEFT)~\cite{Georgi:1990um,Luke:1992cs,Neubert:1993mb,manohar_wise_2000,Damgaard:2019lfh,Brandhuber:2021kpo, Brandhuber:2021bsf, Brandhuber:2022enp} as well as decoupling limit into pure YM amplitudes. BCJ numerators in HEFT have attracted lots of interest recently for their roles in the computation of gravitational amplitudes for black-hole scattering and gravitational waves~\cite{Brandhuber:2021eyq} ({\it c.f.} ~\cite{Kosower:2018adc,Bern:2019nnu,Damour:2019lcq,Bern:2021dqo,DiVecchia:2021bdo,Herrmann:2021tct,Bjerrum-Bohr:2021din,Bjerrum-Bohr:2021wwt,Jakobsen:2021lvp} for some recent works). We will take the heavy-mass limit of YMs amplitude, and (as we have checked up to $n=10$) highly nontrivial cancellations lead to a nice formula for BCJ numerators in HEFT which corresponds to ${\cal P}_{n{-}3}$ (one dimensional lower)!

Furthermore, our results imply new recursion relations and surprisingly, ``factorization'' properties of BCJ numerators on facets of permutohedra; all these can be extended to BCJ numerators of general YMS amplitudes, which in turn combine into a formula for the YM case as well. For the latter, we can then turn the logic around: since the BCJ numerators are manifestly gauge invariant in $n{-}1$ gluons, by showing that all spurious poles indeed cancel in the amplitude based on such ``factorizations", it follows from the uniqueness theorem of~\cite{Arkani-Hamed:2016rak} that they must give correct YM and gravity amplitude (after double copy) even without knowing the CCK duality.  

Let us consider color-ordered YMS amplitude $A(1^\phi,2,\ldots,n{-}1,n^\phi)$ with scalars $1^\phi,n^\phi$. Its expansion onto the Kleiss-Kuij(KK) basis~\cite{Kleiss:1988ne} of bi-adjoint $\phi^3$ amplitudes has  BCJ master numerators as coefficients reads
\begin{equation}\label{eq: BCJnum}
 A(1^\phi,2,\ldots,n{-}1,n^\phi){=}\sum_{\beta\in S_{n-2}} K(1,\beta,n)A^{\phi^3}(1,\beta,n),
\end{equation}
where the sum is over $(n{-}2)!$ permutations of gluons and $A^{\phi^3}(1,\beta,n) \equiv m(1,2,\ldots,n | 1,\beta,n)$ denotes bi-adjoint $\phi^3$ amplitudes with the first ordering fixed to be $(1,2,\ldots,n)$. 
Remarkably, the BCJ numerators from CCK duality $K(1,\beta,n)$ respect the Bose symmetry of all the $n{-}2$ gluons ~\cite{Cheung:2021zvb}: we only need a single numerator with the ordering chosen to be $\beta=(2,\ldots,n{-}1)$, and all others can be obtained by relabelling; they are also gauge invariant for the gluons, which becomes manifest since the dependence on polarizations is through Lorentz products of linearized field strengths $F_i^{\mu\nu}\equiv p_i^\mu \varepsilon_i^\nu-p_i^\nu\varepsilon_i^\mu$
\begin{equation}\label{Fprod}
[F_{\sigma}]^{\mu \nu}= [F_{\sigma_{1}}\cdot F_{\sigma_{2}}\cdots \cdot F_{\sigma_{|\sigma|}}]^{\mu \nu}
\end{equation}
for an ordered subset $\sigma$. The price to pay for these desirable properties is the presence of $2^{n{-}2}-1$ spurious poles, one for each nonempty subset $I\subset \{2, \ldots, n{-}1\}$:
\begin{equation}\label{Ddef}
  D_{I}:= p_{I}\cdot q_{I},\quad  \text{with} \quad p_I:=\sum_{i\in I}p_i, 
\end{equation} 
which depends on a reference momentum  $q_{I}$. These numerators can be simplified with some choices of $q_I$, and the final amplitude is independent of them. 


\section{The permutohedron and algebra underlying BCJ numerators}
In this section, we show that all the terms in a BCJ master numerator obtained from CCK duality for YMS amplitudes are in one-to-one correspondence with all boundaries of permutohedron $\mathcal{P}_{n{-}2}$, or equivalently terms from a quasi-shuffle product. 

\subsection{The (combinatorial) permutohedra and quasi-shuffle products}\label{sec:def}
Following~\cite{Cheung:2021zvb}, we organize $K(1,2,\ldots, n)$ according to the spurious pole structure, which is isomorphic to the boundary structure of the permutohedron $\mathcal{P}_{n-2}$.

The permutohedron $\mathcal{P}_{n-2}$ is an $(n{-}3)$-dimensional polytope~\cite{wiki:Permutohedron}, whose co-dimension $d$ boundary $\Gamma_d$ can be labeled by $d{+}1$ consecutive subsets
\begin{equation}\label{eq: lBD}
    \Gamma_d:=\{I_{0}, I_{1},\ldots, I_{d}\},
\end{equation}
where $I_d{\neq}\emptyset$ and $I_d\subset I_{d{-}1}\subset\ldots\subset I_0=\{2,3,\ldots,n{-}1\}$; the interior of ${\cal P}_{n{-}2}$ can be viewed as its co-dimension $0$ boundary, $\Gamma_0:=I_0$. ${\cal P}_{n{-}2}$ and its boundaries have appeared in the context of cubic tree graphs from the worldsheet~\cite{Gao:2017dek, He:2018pue}. Here each term of the BCJ numerator $K(1, 2, \ldots, n)$ with $d{+}1$ spurious poles corresponds to such a co-dimension $d$ boundary, thus the numerator can be expanded in terms of boundaries of $\mathcal{P}_{n-2}$
\begin{equation}\label{eq: KBDexp}
K(1,2,\ldots,n)=\sum_{d=0}^{n{-}3}\sum_{\Gamma_d\in\partial^d\mathcal{P}_{n{-}2}}K_{\Gamma_d}(1,2,\ldots,n),
\end{equation}
where we sum over all boundaries $\Gamma_d\in\partial^d\mathcal{P}_{n{-}2}$ with co-dimension $d=0, \ldots, n{-}3$, and the contribution from $\Gamma_d$, $K_{\Gamma_d}(1,2,\ldots,n)\equiv K_{\Gamma_d}$ reads 
\begin{equation}\label{eq: KBD}
K_{\Gamma_d}=\prod_{k=0}^{d}\frac{p_{1
\Delta(I_{k},I_{k+1})
}\cdot F_{\tau_{k}}\cdot q_{I_{k}}}{D_{I_{k}}}.
\end{equation}
It has $d{+}1$ factors each with a denominator $D_{I_k}$ of \eqref{Ddef} and a numerator of the form $p_{1 \Delta(I_{k},I_{k+1})}\cdot F_{\tau_k} \cdot q_{I_k}$ for $k=0, \dots, d$. To specify the ordered subset $\tau_k$ of the Lorentz product as in \eqref{Fprod}, we introduce an alternative form of \eqref{eq: lBD} using ordered sets:
\begin{align}\label{eq: lBDt}
   \Gamma_d
   =&\{\tau_0\cup\tau_1\cup\ldots\cup\tau_d,\tau_1\cup\ldots\cup\tau_d,\ldots,\tau_{d{-}1}\cup\tau_d,\tau_d\}\\\nonumber
   \sim&\{\tau_0,\tau_1,\ldots,\tau_d\}, 
\end{align}
where the first line is equivalent to \eqref{eq: lBD} but we use ordered sets $\tau_{k}=\mathrm{Id}(I_k/I_{k{+}1})$ with $I_{d{+}1}\equiv \emptyset$. $\mathrm{Id}(I)$ means sorting the subset $I$ in numerical ordering~\footnote{For $K(1,\beta,n)$, everything stays the same except that in \eqref{eq: lBDt} the definition of $\tau_i$'s become $\tau_{k}=\beta(I_k/I_{k{+}1})$, {\it i.e.} any subset $I$ is sorted according to $\beta$ ordering.}. Perhaps the most subtle point is that we also define 
$\Delta(I_{k},I_{k+1}):=\left(\bar{I}_{k}\right)_{<\tau_{{k},1}}$, which refers to the elements in the set $\bar{I}_{k}=\{2,3,\ldots,n-1\}/I_k$ that are numerically smaller than the first element $\tau_{k,1}$ of the ordered set $\tau_{k}$. 

For example, at $n=5$, we have a boundary $\Gamma_2=\{I_0=\{2,3,4\}, I_1=\{2,4\}, I_2=\{4\}\}$; equivalently, we have $\tau_0=\{3\}, \tau_1=\{2\}, \tau_2=\{4\}$, thus we have a term
\begin{equation}
K_{234,24,4}= \frac{p_1\cdot F_3\cdot q_{234}  p_1\cdot F_2\cdot q_{24} p_{123}\cdot F_4\cdot q_4 }{D_{234}D_{24} D_4  }
\end{equation}
where we have used $\Delta(I_2,\emptyset)=\left.\left(\bar{I_2}\right)\right|_{<4}=\{2,3\}$. A more nontrivial example for the latter is for $n{=}9$, $\Delta(\{4,5,7,8\},\{4,8\})=\left.\{2,3,6\}\right|_{<5}=\{2,3\}$.


On the other hand, the boundaries of $\mathcal{P}_ {n-2}$ have a nice quasi-shuffle product interpretation. The quasi-shuffle product $\star$ can be defined between two arbitrary generators $\left(\sigma_0,\sigma_1,\ldots,\sigma_r\right)$ and $\left(\rho_0,\rho_1,\ldots,\rho_s\right)$, where $\sigma_i$ and $\rho_j$ are sets with arbitrary lengths and can also be prompted to the quasi-shuffle Hopf algebra~\cite{Hoffman, Brandhuber:2021bsf, Brandhuber:2022enp, Chen:2022nei}. We summarize the definitions in appendix \ref{sec:appAG} and here we just use the following result of the quasi-shuffle product $\hat{K}(2,\ldots, n{-}1)\equiv(2)\star(3)\star\ldots\star(n-1)$
\begin{equation}\label{eq: part}
    \hat{K}(2,\ldots,n{-}1)
    {=}\sum_{d{=}0}^{n{-}3}\sum_{\tau\in \text{part}^{(d{+}1)}(2,\ldots,n{-}1)}\!\!\!\!\!\!\!\!\!\!\!\!\!\!(-1)^{n+d-1}(\tau_0,\tau_1,\ldots,\tau_{d}),
\end{equation}
where $\text{part}^{(d+1)}(2,\ldots,n{-}1)$ denotes all the ordered partitions of $\{2,3,\ldots,n{-}1\}$ into $d+1$ nonempty subsets $(\tau_0,\tau_1,\ldots,\tau_d)$ (each $\tau_i$ is sorted according to $\beta$).  Terms on the RHS of \eqref{eq: part} are in one-to-one correspondence with boundaries of $\mathcal{P}_{n-2}$ as in~\eqref{eq: lBDt}, thus we can rewrite \eqref{eq: KBDexp} in terms of the quasi-shuffle product
\begin{equation}
\label{eq:qsp}
K(1,2,\ldots,n)=\langle\hat{K}(2,\ldots n{-}1)\rangle,
\end{equation}
where we have defined a {\it linear map} $\langle\cdot\rangle$ from \eqref{eq: KBD} for any partition $(\tau_0,\tau_1,\ldots,\tau_{d})$ 
\begin{equation}\label{eq:linmap}
    \langle(\tau_0,\tau_1,\ldots,\tau_{d})\rangle{=}(-1)^{n+d-1}\prod_{k=0}^{d}\frac{p_{1\Delta(I_{k},I_{k+1})
}\cdot F_{\tau_{k}}\cdot q_{I_{k}}}{D_{I_{k}}}.
\end{equation}

\subsection{The counting and some examples}
By definition, the permutohedron $\mathcal{P}_m$ contains $m!$ vertices and $2^{m}-2$ co-dimension one facets. More generally, the number of co-dimension $d$ boundaries of this polytope is $(d+1)!S(m,d+1)$, where $S(m,d)$ is the second kind of Stirling number \cite{wiki:Stirling_numbers_of_the_second_kind}.  Algebraically, $S(m,d+1)$ also counts the number of ways to partition a set of $m$ labeled objects into $d+1$ nonempty unlabeled subsets $\{\tau_0,\tau_1,\ldots,\tau_d\}$ \cite{wiki:Stirling_numbers_of_the_second_kind}, so after considering the ordering between these sets, there are $(d+1)!S(m,d+1)$ terms in the summation for any $d$. 
The total number is the Fubini number $\mathcal{F}_{m}$, where $\mathcal{F}_{m}=\sum_{d=1}^{m}d!S(m,d)$ \cite{Mezo}, thus the $n$-point BCJ numerator has $\mathcal{F}_{n-2}$ terms.
\begin{table}[H]
  \centering
    \begin{tabular}{|c|c|c|c|c|c|}
    \hline
    \diagbox[innerwidth=0.5cm]{n}{d}      & 0     & 1     & 2     & 3 &Total \\
    \hline
    3     & 1 &       &       &  &1\\
    \hline
    4     & 1 & 2 &       &  &3\\
    \hline
    5     & 1 & 6 & 6 & &13 \\
    \hline
    6     & 1 & 14 & 36 & 24&75 \\
    \hline
    \end{tabular}%
  \label{tab:BDc}%
  \caption{Counting co-dimension-$d$ boundaries of $\mathcal{P}_{n-2}$}
\end{table}%

Let us illustrate  \eqref{eq: KBDexp} and \eqref{eq: KBD} with some examples. The most trivial case is $n=3$, where the BCJ numerator corresponds to the zero-dimensional permutohedron $\mathcal{P}_{1}$ which is just a point. It contains one term with $\Gamma_0=\{I_0\}=\{2\}$, thus $K(1,2,3)=K_{2}(1,2,3)=\frac{p_1\cdot F_{2}\cdot q_2}{D_2}$, where we have used $\Delta(I_{0},I_{1})=\emptyset$, this is generally true since $\bar{I_0}=\emptyset$.

\begin{figure}[H]
    \centering
    \subfloat[]{ \label{fig:P2}
    \begin{minipage}[c]{1\linewidth}
    \centering
    \includegraphics[scale=1.5]{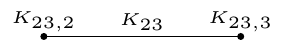}
    \end{minipage}
    }
    
    \subfloat[]{ \label{fig:P3}
    \begin{minipage}[c]{1\linewidth} 
    \centering
    \includegraphics[scale=1.3]{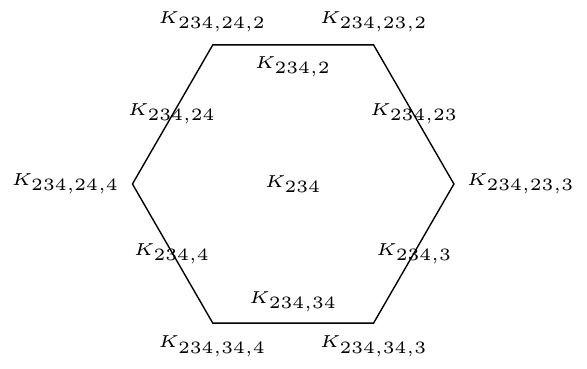}
    \end{minipage}
    }
    \caption{Permutohedra ${\cal P}_2$ for $K(1,2,3,4)$(top) and ${\cal P}_3$ for  $K(1,2,3,4,5)$(bottom)}
\end{figure}

For $n=4$, the permutohedron $\mathcal{P}_{2}$ is a line segment, where the interior ($d=0$) is labelled by $I_0=\{23\}$, and the two vertices ($d=1$) are labeled by $\{23,2\}$ and $\{23,3\}$; we show these three terms in figure \ref{fig:P2}.

Equivalently, in \eqref{eq: part} the partition $\text{part}^{(1)}$ of $\{2,3\}$ has $(\{\tau_0=\{2, 3\})$ and $\text{part}^{(2)}$ has  $(\tau_0=\{2\},\tau_1=\{3\})$ and $(\tau_0=\{3\},\tau_1=\{2\})$: they are nothing but the interior and the two vertices, according to \eqref{eq: lBDt}. 

Thus the BCJ numerator $K(1,2,3,4)$ has three terms, $K_{23}$, $K_{23,2}$ and $K_{23,3}$, which read
\begin{align} \label{eq: BCJnum4}
\frac{p_{1}{\cdot}F_{23}{\cdot} q_{23}}{D_{23}}{+}\frac{p_{1}{\cdot} F_{3}{\cdot} q_{23} p_{1}{\cdot} F_{2}{\cdot} q_{2}}{D_{23}D_{2}}{+}\frac{p_{1}{\cdot} F_{2}{\cdot} q_{23} p_{12}{\cdot} F_{3}{\cdot} q_{3} }{D_{23}D_{3}}
\end{align}
Notice that the last term in the above equation is from the boundary $\{23,3\}$, so the second factor in the numerator is $p_{1\left(\overline{\{3\}}\right)_{<3}}\cdot F_{3}\cdot q_{3}=p_{12}\cdot F_{3}\cdot q_{3}$. Meanwhile \eqref{eq: BCJnum4} shows that the four-point numerator $K(1,2,3,4)$ has an overall pole $D_{I_0}=D_{23}$. This can be easily seen from \eqref{eq: lBD} since the first set of any co-dimension $d$ boundary $\Gamma_d$ is always labeled by $I_0=\{2,3,\ldots,n{-}1\}$.

Notice that \eqref{eq: KBD} means each term in the BCJ numerator contains spurious poles, and the co-dimension $d$ contribution will have $d{+}1$ spurious poles where $D_{I_0}$ is an overall pole for every boundary.
Except for the overall one, the simple poles can be written as $D_I$ where $I$ is a nonempty proper subset of $I_0=\{2,3,\ldots,n{-}1\}$ and two simple poles $D_I$ and $D_J$ are compatible if and only if $I\subset J$ or $J\subset I$. For example, at five-point, except for the overall $D_{234}$, the simple poles are $D_{2},D_{3},D_{4},D_{23},D_{24},D_{34}$ which correspond to the six co-dimension one boundaries of $\mathcal{P}_3$, and the compatible double poles are
\begin{equation}
    \{D_{23}D_{2},D_{24}D_{2},D_{23}D_{3},D_{34}D_{3},D_{24}D_{4},D_{34}D_{4}\},
\end{equation}
which correspond to six vertices of the hexagon $\mathcal{P}_3$. We show the boundary contribution formally in figure \ref{fig:P3}.
These $13$ terms form a two-dimensional polytope $\mathcal{P}_{3}$. These boundaries can also be realized in quasi-shuffle product $\hat{K}(2,3,4)$, which will be discussed in appendix \ref{sec: N5}. To be precise, we give some explicit examples of different co-dimension here
\begin{equation}
\begin{aligned}
    &K_{234}{=}\frac{p_1{\cdot}F_{234}\cdot q_{234}}{D_{234}},\\
    &K_{234,23}{=}\frac{p_1{\cdot}F_4{\cdot}q_{234} p_1{\cdot} F_{23}{\cdot}q_{23}}{D_{234}D_{23}},\;\,K_{234,2}{=}\frac{p_1{\cdot} F_{34}{\cdot} q_{234} p_1{\cdot}F_2{\cdot}q_2}{D_{234}D_2}\\
    &K_{234,23,2}{=}\frac{ p_1{\cdot}F_4{\cdot}q_{234} p_1{\cdot} F_3{\cdot} q_{23} p_1{\cdot} F_2{\cdot} q_2}{D_{234}D_{23}D_2 }.
\end{aligned}
\end{equation}
The complete result for the BCJ numerator $K(1,2,3,4,5)$ is shown in the appendix \ref{sec: N5}.

Moreover, we emphasize that all the spurious poles are canceled in the final amplitude, and the amplitude does not depend on the reference momenta. The proof will be put into the following paper \cite{toapp}.

\begin{figure}[H]
    \centering
    \includegraphics[scale=0.4]{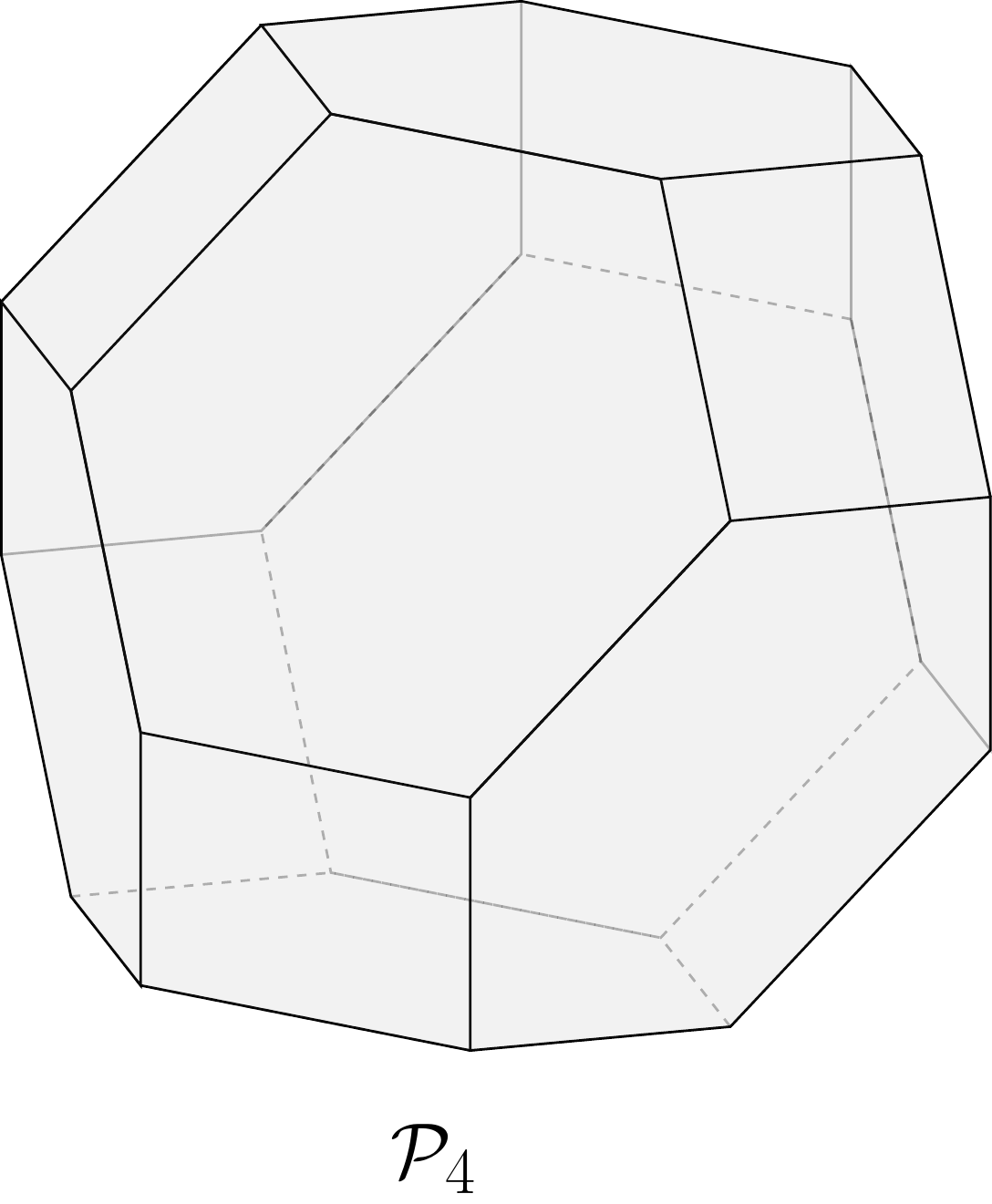}
    \caption{The permutohedron $\mathcal{P}_4$ for $K(1,\ldots, 6)$}
    \label{fig: P4}
\end{figure}

For $n=6$, $\mathcal{P}_4$ is a three-dimensional truncated octahedron shown in figure \ref{fig: P4}. As we have counted, it contains $14$ co-dimension one boundaries (six squares and eight hexagons), $36$ edges and $4!=24$ vertices, thus $75$ boundaries in total. Some terms with co-dimension $d=0,1,2,3$ are
\begin{align}
    &K_{2345}=\frac{p_1{\cdot} F_{2345}{\cdot} q_{2345}}{D_{2345}}\\\nonumber
    &K_{2345,234}=\frac{p_1{\cdot} F_{5}{\cdot} q_{2345}p_1{\cdot} F_{234}{\cdot} q_{234}}{D_{2345}D_{234}}\\\nonumber
    &K_{2345,234,23}=\frac{p_1{\cdot} F_{5}{\cdot} q_{2345}p_1{\cdot} F_{4}{\cdot} q_{234}p_1{\cdot} F_{23}{\cdot} q_{23}}{D_{2345}D_{234}D_{23}}\\\nonumber
    &K_{2345,234,23,2}=\frac{p_1{\cdot} F_{5}{\cdot} q_{2345}p_1{\cdot} F_{4}{\cdot} q_{234}p_1{\cdot} F_{3}{\cdot} q_{23}p_1{\cdot} F_2{\cdot} q_2}{D_{2345}D_{234}D_{23}D_2}
\end{align}
\section{Numerators for general YMS and pure YM amplitudes}
\label{sec:YM}
More generally, the CCK duality has provided closed-formulae for BCJ numerators of $n$-point YMS amplitude with $r\geq 2$ scalars~\cite{Cheung:2021zvb}. It turns out that any such numerator corresponds to a permutohedron ${\cal P}_{n{-}r}$ (with dimension $n{-}r{-}1$): everything we have discussed above for $r=2$ case still applies if we replace $\{2, 3,\ldots, n{-}1\}$ by the set of $n{-}r$ gluons. For example, in the other extreme with $r=n$ we can formally define ${\cal P}_0:={\cal P}_{\emptyset}$ and the $n$-scalar numerator (of bi-adjoint $\phi^3$ amplitude) is $1$ or $0$. For $r=n{-}1$, we have the zero-dimensional ${\cal P}_1$ and the numerator with a single gluon $i$ reads
\begin{equation}
K^{1-{\rm gluon}}(1, \beta, n)=\frac{p_{1{\rightarrow }i} \cdot F_i \cdot q_i}{D_i}
\end{equation} where $1 {\rightarrow} i$ denotes all the scalars preceding $i$ in $(1\beta n)$. We shall not repeat this for general cases but leave detailed discussions to a separate paper~\cite{toapp}. 

Moreover, since the pure YM amplitude can be expanded as a linear combination of these YMS ones~\cite{Lam:2016tlk, Fu:2017uzt, Du:2017kpo, Cheung:2017ems, Dong:2021qai}, we obtain its BCJ numerators for free; the resulting numerator naively contains $2\mathcal{F}_{n-2}$ terms as derived in \cite{Cheung:2021zvb}. However, we can still organize the terms according to pole structures and immediately combine them in pairs as $\mathcal{F}_{n-2}$ terms: the resulting numerator has the same form as the $2{-}$scalar case and corresponds to boundaries of permutohedron $\mathcal{P}_{n-2}$. By expanding $A^{\text{YM}}(1,2,\ldots,n)$ in exactly the same way as \eqref{eq: BCJnum},
each master BCJ numerator, {\it e.g.}  $K^{\text{YM}}(1,2,\ldots,n)$ is given by a sum over boundaries of $\mathcal{P}_{n-2}$ as in \eqref{eq: KBDexp}:
\begin{align}
    K^{\text{YM}}(1,2,\ldots,n)&=\sum_{d=0}^{n{-}3}\sum_{\Gamma_d\in\partial^d\mathcal{P}_{n{-}2}}K^{\text{YM}}_{\Gamma_d}(1,2,\ldots,n). \nonumber
\end{align}
where the contribution from each boundary 
is identical to \eqref{eq: KBD} except for the $k=0$ factor which becomes
\begin{equation}\label{eq: N0YM}
    \frac{\varepsilon_n\cdot F_{1\tau_0}\cdot q_{I_{0}}+\varepsilon_1\cdot F_{\tau_0}\cdot\left(\varepsilon_n p_{1n}\cdot q_{I_{0}}-q_{I_{0}} p_{1}\cdot\varepsilon_n\right)}{D_{I_0}}
\end{equation}
Of course, similar to the YMS case, all spurious poles cancel in the final amplitude, which does not depend on $q_I$. Therefore we are free to choose them to simplify the expression \eqref{eq: N0YM}. One such choice is $q_{I_{0}}=\varepsilon_n$, and the $k=0$ factor \eqref{eq: N0YM} takes a simpler form 
\begin{equation}\label{eq: KYMSim}
    \frac{\varepsilon_n{\cdot} F_{1\tau_0}{\cdot} \varepsilon_n}{\varepsilon_n{\cdot} p_{23\ldots n{-}1}}=-\frac{\varepsilon_n{\cdot} F_{1\tau_0}{\cdot} \varepsilon_n}{\varepsilon_n{\cdot} p_{1}}.
\end{equation}
It is easy to see that the BCJ numerators become manifestly gauge invariant in particles $1,2,\ldots,n{-}1$.
For example, the BCJ numerator $K^{\text{YM}}(1,2,3,4)$ reads
\begin{equation*}
    {-}\frac{\varepsilon_{4}{\cdot}F_{123}{\cdot} \varepsilon_4}{p_1{\cdot}\varepsilon_4}{-}\frac{\varepsilon_4{\cdot} F_{13}{\cdot} \varepsilon_4 p_{1}{\cdot} F_{2}{\cdot} q_{2}}{p_1{\cdot}\varepsilon_4 D_{2}}{-}\frac{\varepsilon_4{\cdot} F_{12}{\cdot} \varepsilon_4 p_{12}{\cdot} F_{3}{\cdot} q_{3} }{p_1{\cdot}\varepsilon_4 D_{3}}.
\end{equation*}

Furthermore, similar to the discussion in section \ref{sec:def}, BCJ numerators of YM amplitudes can also be interpreted in terms of quasi-shuffle products, and the only change is that in the linear map \eqref{eq:linmap} the $k=0$ factor is modified to  \eqref{eq: N0YM}. 

Before ending the section, we mention the obvious double-copy from YM to GR
\begin{equation}
M^{\rm GR}_n=\sum_{\alpha, \beta} K^{\rm YM}(1, \alpha, n) m(1,\alpha ,n|1, \beta ,n) K^{\rm YM}(1, \beta, n)    
\end{equation}
where we sum over a pair of permutations $\alpha,\beta$ of $\{2,3,\ldots, n{-}1\}$, with $m$ denoting bi-adjoint $\phi^3$ amplitudes; if we replace YM by YMS with $1, n$ being scalars, it gives the amplitude with $n{-}2$ gravitons and two scalars. 

\section{Recursions and factorizations}
In this section, we propose recursion relations and factorization properties (on spurious poles $D_I$) for the BCJ numerators, which are implied by the combinatorial and algebraic structure. The argument can be equally applied to both two-scalar YMS and pure YM numerators.

\subsection{Recursion relations}
First, in quasi-shuffle product~\eqref{eq: part}, one can collect the terms with the same $\tau_d$ and then apply the linear map \eqref{eq:linmap} to obtain the following recursion relation,
\begin{equation}\label{eq: Krec}
    K(1,2,\ldots,n){=}
    \sum_{I\subset\{2,\ldots,n{-}1\}}\frac{p_{1\Delta(I,\emptyset)
    }{\cdot} F_I {\cdot} q_I}{D_I}\tilde{K}^I(1,\bar{I},n),
\end{equation}
where the summation is over all the nonempty subsets of $\{2,3,\ldots,n{-}1\}$. The definition of $\tilde{K}^I(1,\bar{I},n)$ is slightly different from \eqref{eq: KBD} in the denominator: it is given by the sum over boundaries of the permutohedron $\mathcal{P}_{\bar{I}}$ with vertices labeled by all permutations of set $\bar{I}$, and for each boundary $\Gamma_d=\{J_0=\bar{I},J_1,\ldots,J_{d}\}$ where $J_d\subset J_{d{-}1}\ldots\subset J_0$, we have a  contribution 
\begin{equation}\label{eq: tKBD}
\tilde{K}^I_{\Gamma_d}=\prod_{k=0}^{d}\frac{p_{1\Delta(J_{k},J_{k+1})
}\cdot F_{\tau_k}\cdot q_{IJ_{k}}}{D_{IJ_{k}}},
\end{equation}
where $D_{IJ_k}=p_{IJ_k}\cdot q_{IJ_k}$, $\tau_k= \mathrm{Id}(J_k/J_{k{+}1})$ with $J_{d{+}1}\equiv \emptyset$ and the complement of the set $J_k$ appears in $\Delta(J_{k},J_{k+1})$ is defined as $\bar{I}/J_k$. For $\abs{I}{=}n{-}2$ ($\bar{I}=\emptyset$), we define $\tilde{K}^I(1,n)=1$. Formally, this numerator corresponds to the permutohedron $\mathcal{P}_0$.

\begin{figure}[H]
    \centering
    \includegraphics[scale=0.8]{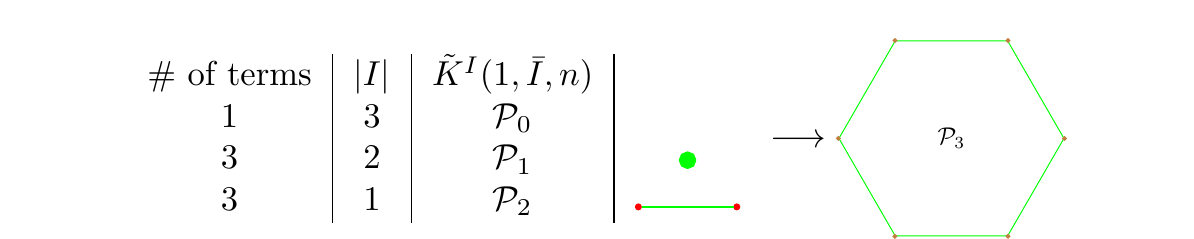}
    \caption{Recursion relation at $n=5$}
    \label{fig:rec5}
\end{figure}

For example, the recursion relation of $K(2,3,4)\equiv K(1,2,3,4,5)$ \footnote{Here we have omitted the labels of the scalar particles $1$ and $n$ in the numerators $K$ and $\tilde{K}$.} reads,
\begin{small}
\begin{align}\label{eq: rec5}
    K&(2,3,4)=\frac{p_1{{\cdot}} F_{234}{\cdot} q_{234}}{D_{234}}\\\nonumber
    {+}&\frac{p_1{\cdot} F_{23}{\cdot} q_{23}}{D_{23}}\tilde{K}^{23}(4){+}\frac{p_{1}{\cdot} F_{24}{\cdot} q_{24}}{D_{24}}\tilde{K}^{24}(3){+}\frac{p_{12}{\cdot} F_{34}{\cdot} q_{34}}{D_{34}}\tilde{K}^{34}(2)\\\nonumber
    {+}&\frac{p_1{\cdot} F_2{\cdot} q_2}{D_2}\tilde{K}^2(3,4)
    {+}\frac{p_{12}{\cdot} F_3{\cdot} q_3}{D_3}\tilde{K}^3(2,4){+}\frac{p_{123}{\cdot} F_4{\cdot} q_4}{D_4}\tilde{K}^4(2,3).
\end{align}
\end{small}
Geometrically, the recursion relation~\eqref{eq: Krec} tells us how the co-dimension one boundaries of permutohedron are glued together. In the above five-point example, the term with $\abs{I}=3$ in the first line has only one pole and corresponds to the interior (co-dimension $0$ boundary) of $\mathcal{P}_{3}$, depicted in figure~\ref{fig:rec5}. For the three terms with $\abs{I}=2$ in the second line, each factor $\tilde{K}^I(1,\bar{I},n)$ corresponds to a zero-dimensional permutohedron; on the other hand, each term is mapped to a co-dimension one boundary of $\mathcal{P}_3$ without vertices. For the remaining three terms with $\abs{I}=1$ in the last line, each $\tilde{K}^I(1,\bar{I},n)$ corresponds to a one-dimensional permutohedron and it is mapped to a co-dimension one boundary with two vertices.
\subsection{Factorization properties on spurious poles}
Next, we move to certain intriguing factorization properties of the BCJ numerator on spurious poles. Combinatorially, any co-dimension one boundary of the permutohedron $\mathcal{P}_{n{-}2}$ is the product of two lower-dimensional permutohedra $\mathcal{P}_{I}\times\mathcal{P}_{\bar{I}}$. Remarkably, we find that on any pole $D_I=0$, the residue of the BCJ numerator factorizes into the product of a $(\abs{I}{+}2)$-point numerator and a $(n{-}\abs{I})$-point numerator! Unlike the usual factorization on the physical poles of the amplitude, these factorizations on the spurious poles stem from the combinatorial picture without any known physical origin. Explicitly
\begin{equation}\label{eq: Kfac}
\begin{aligned}
\left.\mathrm{Res}\right|_{D_I{=}0}K(1,2,\ldots,n){=}&D_I K( 1,\mathrm{Id}(I),P) \\
    &\times\tilde{K}^I(1,\mathrm{Id}(\bar{I}),n),
\end{aligned}
\end{equation}
where $P\equiv \bar{I}n$ denotes an effective scalar. For the definition of $\Delta(I_{k}, I_{k{+}1})$ in $K( 1,\mathrm{Id}(I),P)$, the complement of the set $I_k$ is still defined as $\bar{I_k}=\{2,3,\ldots,n{-}1\}/I_k$ while for $\Delta(J_{k}, J_{k{+}1})$ in $\tilde{K}^I(1,\mathrm{Id}(\bar{I}),n)$ the complement of $J_k$ is defined as $\bar{I}/J_k$. 
The factor $D_I K(1,\mathrm{Id}(I),P)$ in \eqref{eq: Kfac} means that the overall pole $D_I$ of $K(1,\mathrm{Id}(I),P)$ is excluded. The factorization properties~\eqref{eq: Kfac} can be proved directly by plugging in the definitions on both sides.

For instance, at six points as shown in figure \ref{fig: P4}, there are $14$ co-dimension one boundaries $D_I=0$ including eight poles with $\abs{I}=1$ or $3$ corresponding to hexagons and six poles with $\abs{I}=2$ corresponding to squares. On any of the hexagon boundary, {\it i.e.} when $D_I=0$ with $\abs{I}=1$ or $3$, the residue factorizes into $\mathcal{F}_3=13$ terms (times $\mathcal{F}_1=1$ term). Similarly when $D_I=0$ with $\abs{I}=2$, the residue factorizes differently, {\it e.g.} as $D_{23}K(1,2,3,456)\times\tilde{K}^{23}(1,4,5,6)$ when $D_{23}=0$ (the square is the product of two line segments $\mathcal{P}_{\{23\}}\times\mathcal{P}_{\{45\}}$). 

Algebraically, the quasi-shuffle algebra can be prompted to a bialgebra by introducing the coproduct map \cite{Hoffman}, and one can show the factorization properties from the coproduct. Similar to \cite{Brandhuber:2021bsf, Brandhuber:2022enp}, we can also define the antipode map to make the bialgebra a quasi-shuffle Hopf algebra. Acting on the BCJ numerators, the antipode map does nothing but changes its overall sign. The detail is given in the appendix \ref{sec:appAG}. 

We expect the factorization properties of BCJ numerators to be the key for showing the cancellation of spurious poles in the amplitude. Such properties also suggest certain positive geometries (rather than just combinatorics) underlying these BCJ numerators, and we leave further investigations to future works. 


\section{Heavy-mass effective field theory} \label{sec:heavy1}
In this section, we study YMS amplitudes and their BCJ numerators in the heavy-mass effective field theory (HEFT), which are obtained by taking the heavy-mass limit for a pair of massive scalars with momenta \cite{Brandhuber:2021kpo, Brandhuber:2021bsf,Brandhuber:2022enp} 
\begin{equation}
    p_1^\mu=mv^\mu,\qquad p_n^\mu=-m v^\mu - k^\mu, 
\end{equation}
where $v^2=1$ and we are interested in the limit $m \to \infty$; in other words, we will study the the expansion in $1/m$ of the BCJ numerators which we denote as $K_\mathrm{H}(1,2,\ldots,n)$, as well as that of $\phi^3$ amplitudes, which combine to give the resulting HEFT amplitude $A^\mathrm{H}(1,2,\ldots,n)$ at the leading order in $1/m$. Here $k^\mu$ is at the same order as gluon momenta, which stay finite at ${\cal O}(m^0)$ as $m \to \infty$. 

\subsection{Heavy limit of YMS amplitudes}
We will make a  particular choice of the reference momenta: $q_I=v$ for all $I$, which dramatically simplifies formulae for BCJ numerators and give rise to poles similar to HEFT numerators  in~\cite{Brandhuber:2021bsf}. In fact, for $n=4$ such a choice reduces the BCJ numerator to one term, since $v\cdot F_a \cdot v$ vanishes for a single particle $a$
\begin{equation}\label{eq: hamp4}
    K_\mathrm{H}(1,2,3,4)= \frac{p_{1}\cdot F_{23}\cdot v} {p_{23} \cdot v} = -\frac{2m^2}{k^2}\ v \cdot F_{23}\cdot v,
\end{equation}
where in the second equality we have used $v\cdot k=-k^2/(2m)$ implied by the on-shell condition $p_n^2=m^2$.
Notice that $K_{\mathrm{H}}(1,3,2,4)=K_{\mathrm{H}}(1,2,3,4)$, thus the amplitude  $A^\mathrm{H}(1,2,3,4)$ becomes
\begin{align} \label{eq:Hamp4}
&(\frac{1}{s_{12}}+ \frac{1}{s_{23}}) K_\mathrm{H}(1,2,3,4)  -\frac{1}{s_{23}}K_\mathrm{H}(1,3,2,4)  \\\nonumber
=&-\frac{m}{k^2} \frac{v \cdot F_{23}\cdot v}{ v\cdot p_2}.
\end{align}
Physically, the final HEFT amplitude has the leading order $\mathcal{O}(m)$ \cite{Brandhuber:2021kpo}. In the above example, we can see that the numerators are at $\mathcal{O}(m^2)$, and the sum of the leading contribution of $\phi^3$ amplitudes at $\mathcal{O}(m^0)$, say $1/s_{23}$ times the corresponding numerators vanishes. Therefore, the sum of the contribution of $\phi^3$ amplitudes at the next order, {\it i.e.} $1/s_{12}$ from $A^{\phi^3}(1,2,3,4)$ times the numerator produces the HEFT amplitude as the first non-vanishing order. This is also the case for $n=5$. 
However, for higher $n$, the numerator contains some additional terms with higher power of $m$. 
To obtain the leading order contribution of HEFT final amplitude, we expand the numerators and the $\phi^3$ amplitudes in $m^{-1}$. Note the overall pole $D_{23\ldots n-1}=v\cdot k$ for BCJ numerators is proportional to $m^{-1}$, we first collect the numerator according to its {\it superficial order} of $m^{-1}$, {\it i.e.} terms with $(i-1)$ $p_1$'s in the numerator, 
\begin{equation} \label{eq: Kexpand}
    K_{\mathrm{H}}(1,2,\ldots,n)= \sum_{i=2}^{\lfloor n/2 \rfloor } K_\mathrm{H}^{(i)}(1,2,\ldots,n),
\end{equation}
where the upper bound of the summation is $\lfloor n/2 \rfloor$ since $p_1 \cdot F_{a}\cdot v=0$ 
implies that the numerator should contain as many $p_1 \cdot F_{ab} \cdot v$ as possible to have the highest power of $p_1$. In the above expansion, $K_\mathrm{H}^{(i)}\equiv K_\mathrm{H}^{(i)}(1,2,\ldots,n)$ refers to terms with the superficial order $\mathcal{O}(m^i)$. For example, at six points we have the following terms for $K_\mathrm{H}^{(2)}$ and $K_\mathrm{H}^{(3)}$ respectively:
\begin{equation*}
       \frac{p_1\cdot F_{23} \cdot v\ p_{23}\cdot F_5 \cdot v\ p_{23}\cdot F_4 \cdot v } {v\cdot k\ v \cdot p_{45}\ v\cdot p_4 }, \frac{p_1 \cdot F_{25}\cdot v\ p_1\cdot F_{34}\cdot v} {v\cdot k\ v\cdot p_{34} }.
\end{equation*}
In fact, as explained in appendix \ref{sec:heavy}, the actual order of $K_\mathrm{H}^{(i)}$ is $\mathcal{O}(m^{2})$ for $i=2$ and $\mathcal{O}(m^{i-1})$ otherwise.

In the HEFT amplitude, we sum over all cubic graphs relevant at leading order, and for each graph with its propagator structure, its numerator is given by the corresponding commutator of $K_\mathrm{H}^{(i)}$~\cite{Bern:2011ia}. Nicely we observe that certain commutators of $K_\mathrm{H}^{(i)}$ actually vanish, and the end result is that only $K_\mathrm{H}^{(2)}$ contributes to the amplitude at the leading order! We have checked such vanishing results up to $n=10$, but we do not have an all-$n$ proof at the moment. 

In fact, such vanishing results are better than what we need here, {\it i.e.} only $K^{(2)}_{\rm H}$ contributes to gauge-theory amplitudes at leading order. We have checked up to $n=10$ that the stronger vanishing results actually ensure that only $K^{(2)}_{\rm H}$ contributes to gravity amplitude, which is at order $\mathcal{O}(m^2)$, as obtained by double copy in HEFT. We leave more details in the appendix \ref{sec:heavy} with a proof of the simplest case. As a result of this conjecture, the amplitude is given by
\begin{equation} \label{eq: Hamp}
\begin{aligned}
    &A^\mathrm{H}(1,2 \ldots,n)=\sum_{\Theta^1 } \frac{K_\mathrm{H}^{(2)}(1,\Theta^1,n)}{d_{\Theta^1}}, \\
\end{aligned}
\end{equation}
where the summation is over nested commutators of depth $n{-}4$ (``co-depth" $1$) of the ordered set $(2,3,\ldots,n-1)$. For instance, at 5-point we sum over $\Theta^1=([2,3],4),(2,[3,4])$; $d_{\Theta^1}$ denotes the propagator denominator corresponding to the cubic tree associated with $\Theta^1$ (two sub-trees on the scalar line $(1 n)$):
\begin{equation*}
\begin{aligned}\includegraphics[width=0.23\linewidth]{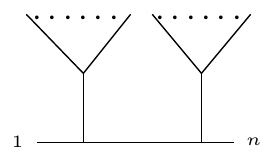}\end{aligned} \leftrightarrow d_{\Theta^{1}}, {\it e.g.} \begin{aligned}\includegraphics[width=0.23\linewidth]{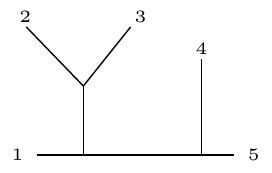}\end{aligned} \leftrightarrow  d_{[2,3],4}=s_{123}s_{23}.
\end{equation*}
Moreover, it is easy to show (see appendix \ref{sec:heavy} for details) that the effective BCJ numerator $K_\mathrm{H}^{(2)}(1,2,\ldots,n)$ contains $\mathcal{F}_{n{-}3}$ terms, and its pole structure corresponds to the permutohedron $\mathcal{P}_{\{34\ldots n{-}1\}}$, which means that
\begin{equation}\label{eq: BDexph}
    K_\mathrm{H}^{(2)}(1,2,\ldots,n)=\sum_{d=0}^{n-2}\sum_{\Gamma_d\in\partial^d\mathcal{P}_{n{-}3}}K^{ (2)}_{H,\Gamma_d}(1,2,\ldots,n).
\end{equation}
For the boundary $\Gamma_d{=}\{I_0,I_1,\ldots,I_d\}{\in}\partial^d\mathcal{P}_{\{34\ldots n{-}1\}}$ where $I_d{\subset}I_{d-1}{\subset}\ldots{\subset}I_0{=}\{3,4,\ldots,n{-}1\}$ and $I_d{\neq}\emptyset$, the contribution is
\begin{align}
    K^{(2)}_{\mathrm{H},\Gamma_d}&= \frac{m v\cdot F_{\tau_{0}}\cdot v}{p_{23\ldots n{-}1}\cdot v} \prod_{k=1}^{d}\frac{p_{\Delta(I_{k},I_{k{+}1})}\cdot F_{\tau_d}\cdot v}{v\cdot p_{I_k}}\\\nonumber
    &=-\frac{2m^2 v\cdot F_{\tau_{0}}\cdot v}{k^2} \prod_{k=1}^{d}\frac{p_{\Delta(I_{k},I_{k{+}1})}\cdot F_{\tau_d}\cdot v}{v\cdot p_{I_k}},
\end{align}
where in the calculation of $\Delta(I_{k},I_{k{+}1})$, the complement set of $I_k$ is still taken to be $\{2,3,\ldots,n{-}1\}/I_k$. For $n=4$, there is no commutator in $\Theta^1$ and the result is \eqref{eq:Hamp4}. 
For $n=5$, the amplitude becomes
\begin{equation}
\begin{aligned}
        A^\mathrm{H}(1,2,3,4,5){=} &\frac{1}{s_{12}} \frac{K_\mathrm{H}^{(2)}(1,2,[3,4],5)}{s_{34}}  \\
        & {+}\frac{1}{s_{123}} \frac{K_\mathrm{H}^{(2)}(1,[2,3],4,5)}{s_{23}},
\end{aligned}
\end{equation}
where $K_\mathrm{H}^{(2)}(1,2,3,4,5)$ is given by{\small
\begin{equation*}
 -\frac{2m^2}{k^2}(v \cdot F_{234} \cdot v+\frac{ v\cdot F_{24} \cdot v\ p_{2}\cdot F_3\cdot v }{v\cdot p_3 }
     + \frac{ v\cdot F_{23} \cdot v \ p_{23}\cdot F_4\cdot v }{v\cdot p_4}).
\end{equation*}}
Let us give a final example for $n=6$ amplitude
\begin{equation}
\begin{aligned}
    & \frac{1}{s_{12}} \left(\frac{K_\mathrm{H}^{(2)}(1,2,[[3,4],5],6)}{s_{34}s_{345}}{+} \frac{K_\mathrm{H}^{(2)}(1,2,[3,[4,5]],6)}{s_{45}s_{345}} \right) \\
    {+}& \frac{1}{s_{123}} \frac{K_\mathrm{H}^{(2)}(1,[2,3],[4,5],6)}{s_{23}s_{45}}  \\
    {+}&\frac{1}{s_{1234}} \left(\frac{K_\mathrm{H}^{(2)}(1,[[2,3],4],5,6)}{s_{23}s_{234}}{+} \frac{K_\mathrm{H}^{(2)}(1,[2,[3,4]],5,6)}{s_{34}s_{234}} \right).
\end{aligned}
\end{equation}
It is interesting to notice the numerators we present here only differ from those in \cite{Brandhuber:2021bsf} denoted by $N(1,2,\ldots,n)$ by an overall prefactor
\begin{equation}\label{eq: HBCJnumerator}
    K_\mathrm{H}^{(2)}(1,2,\ldots,n)=(-1)^{n}(n{-}2)\frac{2m}{k^2} v\cdot p_2 N(1,2,\ldots,n).
\end{equation}
It is highly nontrivial, however, that these two sets of effective BCJ numerators give the same HEFT amplitude. In~\cite{Brandhuber:2021bsf}, the expression involves the sum of cubic graphs corresponding to nested commutators of depth $n-3$ of the ordered set $(2,3,\ldots,n{-}1)$, thus the propagator denominator contains an overall factor $s_{23\ldots n-1}$, which in our case is replaced by different $s_{1 \sigma}$ for different terms. In addition, the numerator of~\cite{Brandhuber:2021bsf} for each cubic graph is given by a nested commutator of $N(1,2,\ldots,n)$, thus the number of terms in it is twice as ours. Nevertheless, we have analytically checked up to $n=10$ that the amplitude \eqref{eq: Hamp} agrees with \cite{Brandhuber:2021bsf}. Moreover, we have checked that although they look very different, the HEFT gravity amplitude via double copy also agrees with that in~\cite{Brandhuber:2021bsf}, and we expect both agreements to hold for all $n$.

\subsection{Decoupling into pure YM}
Given the explicit result of the $n$-point heavy mass BCJ numerators, the $(n-1)$-point pure YM BCJ numerators, as well as the amplitudes, can be easily obtained via the decoupling limit: 
$mv\to \varepsilon_n$, $p_{23\ldots n-1}^2\to 0$ 
to obtain the BCJ numerator $K^{\prime \text{YM}}(2,3,\ldots,n)$ \cite{Brandhuber:2021kpo,Brandhuber:2021bsf}. Under this kinematics, the overall factor $k^2$ vanishes, which we ignore in the decoupling limit. For instance, the three-point BCJ numerator is given by $K^{\prime \text{YM}}(2,3,4)=-2\varepsilon_4 \cdot F_{23}\cdot \varepsilon_4$. Therefore, the three-point amplitude is
\begin{align}
    A^{\text{YM}}(2,3,4)&=  -\frac{\varepsilon_4 \cdot F_{23}\cdot \varepsilon_4}{ \varepsilon_4 \cdot p_2}\\\nonumber
    &=\varepsilon _4\cdot \varepsilon _2 p_2\cdot \varepsilon _3{-}\varepsilon _2\cdot \varepsilon _3 p_2\cdot \varepsilon _4{-}\varepsilon _4\cdot \varepsilon _3 p_3\cdot \varepsilon _2.
\end{align}
For the 4-point YM amplitude, the numerator $K^{\prime \text{YM}}(2,3,4,5)$ reads
\begin{equation*}
    {-}2\left(\varepsilon_{5}{\cdot}F_{234}{\cdot} \varepsilon_5{+}\frac{\varepsilon_5{\cdot} F_{24}{\cdot} \varepsilon_5 p_{2}{\cdot} F_{3}{\cdot} \varepsilon_5}{p_3{\cdot}\varepsilon_5 }{+}\frac{\varepsilon_5{\cdot} F_{23}{\cdot} \varepsilon_5 p_{23}{\cdot} F_4{\cdot} \varepsilon_5 }{ p_4{\cdot}\varepsilon_5}\right).
\end{equation*}
Note that $K^{\prime \text{YM}}(2,3,\ldots,n)$ also manifests the gauge invariance of particles $2,3,\ldots,n-1$. Moreover, it is related to the BCJ numerator given in sec.\ref{sec:YM} via 
\begin{equation*}
    K^{\prime \text{YM}}(2,3,\ldots,n)=2\left. \varepsilon_n \cdot p_2 K^{\text{YM}}(2,3,\ldots,n)\right|_{q_I\to \varepsilon_n}.
\end{equation*}
These numerators, accompanied by different $\phi^3$ amplitudes, produce the same YM amplitude. 


\section{Conclusions and outlook}

In this note, we established a correspondence between BCJ numerators from covariant color-kinematics duality and the combinatorial permutohedra, which are closely related to the quasi-shuffle Hopf algebra. This apply to all YMS amplitudes, but the most interesting case is that with two scalars, whose numerators share the same combinatorial structure as the pure YM ones: each term is mapped to a boundary of $\mathcal{P}_{n-2}$; the contribution from each boundary is almost identical in these two cases, except that we need to modify one factor to take into account the remaining two gluons. We also found nice recursion relations and factorization properties implied by this picture. Finally, based on highly nontrivial cancellations which are needed for both YMS and gravity amplitudes (via double copy) in HEFT, we conjectured a compact formula for their effective numerators; they become closely related to permutohedra $\mathcal{P}_{n-3}$, which, while producing the same amplitude, differ by an overall factor from the numerators in ~\cite{Brandhuber:2021bsf, Brandhuber:2022enp}.

There are numerous open questions for further investigations. First, as we will present in \cite{toapp}, it is interesting to see how lower-dimensional permutohedra for general YMS numerators combine into $\mathcal{P}_{n-2}$ which corresponds to the pure YM ones; we also find interesting combinatorial structures underlying BCJ numerators of amplitudes in NLSM {\it etc.}. Moreover, the somewhat miraculous cancellations that simplify these numerators in HEFT still remain to be proven, which would also be important to establish the correct double copy in HEFT. Since the final amplitudes are independent of reference momenta, all the spurious poles must cancel, which still calls for a direct understanding (without relying on the CCK duality); such an understanding could connect this combinatorial picture (especially the factorizations) to the uniqueness theorem for YM amplitude~\cite{Arkani-Hamed:2016rak, Rodina:2016jyz} and YMS ones via the universal expansion~\cite{Dong:2021qai}. Last but not least, it is tempting to ask: could we combine the permutohedra for BCJ numerators with the associahedra for bi-adjoint $\phi^3$ amplitudes, and obtain a unified geometric understanding of gluon and graviton scattering?

\begin{acknowledgments}
We thank Linghui Hou, Guanda Lin, Tianheng Wang for discussions and collaborations on related projects. This research is supported in part by the Key Research Program of CAS, Grant No. XDPB15 and National
Natural Science Foundation of China under Grant No.
11935013,11947301,12047502,12047503
\end{acknowledgments}

\newpage

\widetext
\appendix

\section{Review of the permutohedra and Hopf algebras}\label{sec:appAG}

\begin{figure}[H]
    \centering
    \begin{subfigure}[c]{0.4\linewidth}
    \centering
    \includegraphics[scale=0.7]{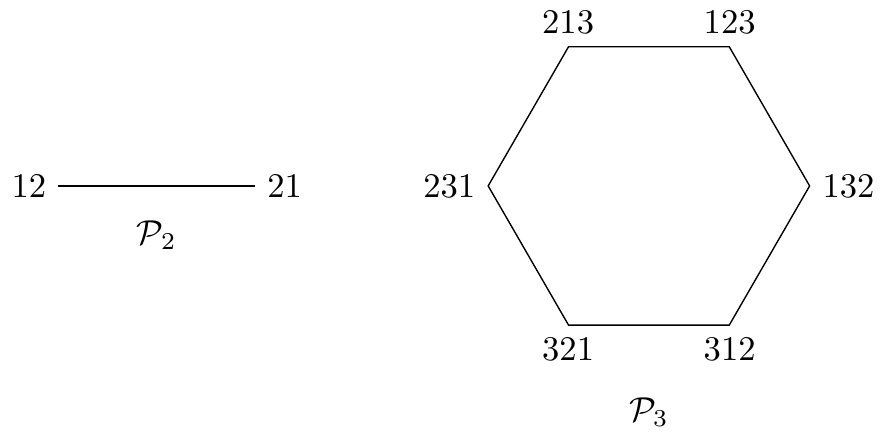}
    \end{subfigure}
    \begin{subfigure}[c]{0.4\linewidth}
    \centering
    \includegraphics[scale=0.5]{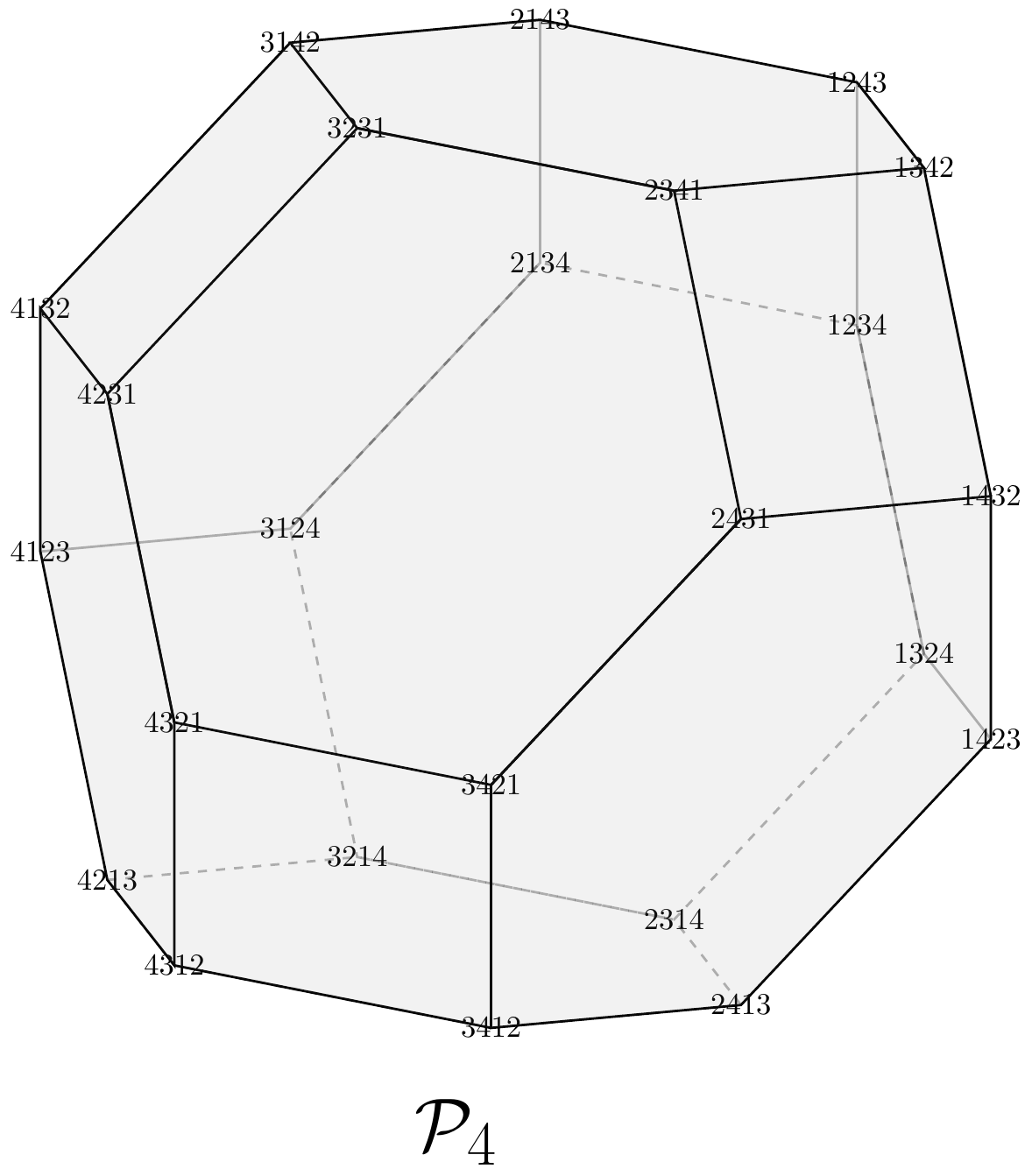}
    \end{subfigure}
    \caption{permutohedra $\mathcal{P}_2$,$\mathcal{P}_3$ and $\mathcal{P}_4$}
    \label{fig:p234}
\end{figure}

\paragraph{The permutohedra} The permutohedron $\mathcal{P}_m$ refers to an $(m-1)$-dimensional polytope, whose vertices are labeled by the $m!$ permutations of $(1,2,3,\ldots,m)$. Two permutations are connected by an edge if and only if they differ in only two places, and the numbers on these places are neighbors \cite{wiki:Stirling_numbers_of_the_second_kind}. For example, the $\mathcal{P}_1$ is just a point, and the $\mathcal{P}_2$ is just a line whose vertices can be labeled as $\{12\}$ and $\{21\}$. For $m=3$, the permutohedron is a hexagon and $\mathcal{P}_4$ is a truncated octahedron, as shown in figure \ref{fig:p234}.

\begin{figure}[H]
    \centering
    \includegraphics{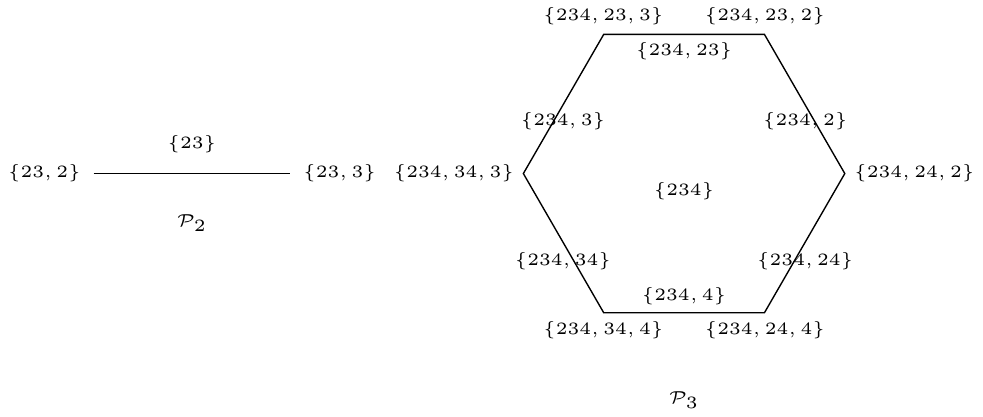}
    \caption{$\mathcal{P}_2$ and $\mathcal{P}_3$ in our convention}
    \label{fig:p23v2}
\end{figure}

It's also easy to translate the traditional label of vertices of permutohedron to our convention. In our problem, the vertices of $\mathcal{P}_{n-2}$ should correspond to the permutations of $\{2,3,\ldots,n-1\}$ denoted by $p=p_1p_2\ldots p_{n-2}$. For any vertex $p$, we apart it into a set $\{\mathrm{Id}(p_1,p_2,\ldots,p_{n-2}), \mathrm{Id}(p_{1},p_{2},\ldots p_{n-3}),\ldots, p_1\}$. Then each co-dimension $n{-}2$-boundary is labeled by the intersection of two vertices. For example, for the $\mathcal{P}_{2}$ as shown in the left side of \ref{fig:p23v2}, the two vertices are now labeled by $\{23,2\}$ and $\{23,3\}$, and the co-dimension $0$ line is just  $\{23,2\}\cap\{23,3\}{=}\{23\}$. More generally, the co-dimension $d$ boundary can be labeled by the intersection of two co-dimension $d{+}1$ boundaries, which is labeled by $d{+}1$ sets. The slightly nontrivial example $\mathcal{P}_3$ is also shown in \ref{fig:p23v2}.

In this new notation, the co-dimension $0$ boundary is always labeled by $I_0\equiv\{2,3,\ldots, n-1\}$ and the co-dimension one boundaries are labeled as $\{I_0,I_1\}$, where $I_1$ is a nonempty proper subset of $I_0$, so there are $2^{n-2}-2$ codimension one boundaries in total. For example, for the permutohedron $\mathcal{P}_4$, the co-dimension one boundaries include eight hexagons (with $\abs{I_1}=1$ or $3$) and six squares (with $\abs{I_1}=2$). Two co-dimension one boundaries $\{I_0, I_1\}$ and $\{I_0, I_1^\prime\}$ are adjacent if and only if the $I_1\subset I_1^\prime$ or $I_1\supset I_1^\prime$, As a co-dimension two boundary, the intersection of the above two boundaries is just $\{I_0, I_1, I_1^\prime/I_1\}$ if $I_1\subset I_1^\prime$ and similar if $I_1\supset I_1^\prime$. It's easy to notice that the boundary structure of $\mathcal{P}_{n-2}$ is just the same as the pole structure in our $n$-point BCJ numerator.



Combinatorially, each co-dimension one boundary of the permutohedron $\mathcal{P}_{n-2}$, say $\{I_0,I_1\}$ is the product of two permutohedra $\mathcal{P}_{I_1}\times\mathcal{P}_{\bar{I_1}}$, which we refer to as ``combinatorial factorization". For example, on a co-dimension one boundary of $\mathcal{P}_4$, say $\{2345,2\}$, the permutohedron factorizes into $\mathcal{P}_{\{2\}}\times\mathcal{P}_{\{345\}}$, which is just a hexagon; on another co-dimension one boundary $\{2345,23\}$, it  factorizes into $\mathcal{P}_{\{23\}}\times\mathcal{P}_{\{45\}}$, {\it i.e.} the product of two line segment, which is a square. Let us also list all the co-dimension one boundaries of $\mathcal{P}_5$. Denoted as $\{23456, I_1\}$, when $\abs{I_1}=1$ or $4$, the boundary is just the truncated octahedron $\mathcal{P}_4$; when $\abs{I_1}=2$ or $3$, the boundary becomes a line segment times a hexagon which is a hexagonal prism. More generally, we can see that boundaries corresponding to $\abs{I_1}=a$ and $\abs{I_1}=n-2-a$ have the same shape.

\paragraph{Hopf algebras}The quasi-shuffle algebra consists of a vector space $V$ of generators, which are denoted as $\left(\sigma_0,\sigma_1,\ldots,\sigma_r\right)$. To begin, we introduce some standard nomenclature for generators: we will refer to
generators with a single subset $(\sigma_0)$ as “letters”, those with multiple subsets $\sigma\equiv\left(\sigma_0,\sigma_1,\ldots,\sigma_r\right)$
as “words”. The quasi-shuffle product between two generators can be defined recursively as
\begin{align}\label{eq: qsrec}
    (\sigma_0,\sigma_1,\ldots,\sigma_r)\star(\rho_0,\rho_1,\ldots,\rho_s)
    =&\sigma_0\left[(\sigma_1,\ldots,\sigma_r)\star(\rho_0,\rho_1,\ldots,\rho_s)\right]+\rho_0\left[(\sigma_0,\sigma_1,\ldots,\sigma_r)\star(\rho_1,\ldots,\rho_s)\right]\\\nonumber
    -&(\sigma_0\rho_0)\left[(\sigma_1,\ldots,\sigma_r)\star(\rho_1,\ldots,\rho_s)\right],
\end{align}
where we also defined an identity element $\mathbb{I}$ for the quasi-shuffle product
\begin{align}
\mathbb{I} \left(\sigma_0\right) \cdots\left(\sigma_r\right) &=\left(\sigma_0\right) \cdots\left(\sigma_r\right) \mathbb{I} =\left(\sigma_0\right) \cdots\left(\sigma_r\right), \\\nonumber
\mathbb{I} \star\left(\sigma_0\right) \cdots\left(\sigma_r\right) &=\left(\sigma_0\right) \cdots\left(\sigma_r\right) \star \mathbb{I} =\left(\sigma_0\right) \cdots\left(\sigma_r\right).
\end{align}

For example, the quasi-shuffle product between two letters is
\begin{equation}\label{eq:shufle2}
    (2)\star(3)=(2,3)+(3,2)-(23),
\end{equation}
and the product between a letter with a word is

\begin{align}
    (2,3)\star(4)&{=}(2) [(3)\star(4)]{+}(4,23){-}(24,3) \\
    &{=}(2,3,4){+}(2,4,3){-}(2,34){+}(4,23){-}(24,3) \nonumber
\end{align}

From \eqref{eq: qsrec}, the quasi-shuffle product between two generators can also be written out explicitly
\begin{equation}
    (\sigma_0,\sigma_1,\ldots,\sigma_r)\star(\rho_0,\rho_1,\ldots,\rho_s)=\!\!\!\!\!\!\!\!\!\!\sum_{\substack{\left.\tau\right|_{\{\sigma\}}=(\sigma_0,\sigma_1,\ldots,\sigma_r)\\\left.\tau\right|_{\{\rho\}}{=}(\rho_0,\rho_1,\ldots,\rho_s)}}\!\!\!\!\!\!\!\!({-}1)^{d{-}r{-}s}(\tau_0,\tau_1,\ldots,\tau_d)
\end{equation}
where the $\sigma_i$ or $\rho_i$ are now any subsets of $\{2,3,\ldots,n{-}1\}$. The notation $\left.\tau\right|_{\{\sigma\}}$ means that we restrict the partition $\tau$ onto the subset $\{\sigma\}=\cup_{i=0}^r\sigma_i$, for example $\left.(234,56,78)\right|_{\{2,4,6\}}=(24,6)$. 

Here, we give the $n=4$ example for the BCJ numerators from the linear map of the quasi-shuffle product~\eqref{eq:qsp}. The BCJ numerators $K(1,2,3,4)=\langle\hat{K}(2,3)\rangle$ can be mapped from the quasi-shuffle product~\eqref{eq:shufle2} via the rule~\eqref{eq:linmap}, which is the same as the result shown in~\eqref{eq: BCJnum4}.

As we mentioned in the main text, the quasi-shuffle product $(2)\star(3)\star\ldots\star(n-1)$ gives the sum over all the ordered partitions of $\{2,3,\ldots,n-1\}$ into $d+1$ nonempty subsets with $d=0,1,\ldots,n-3$ (see \eqref{eq: part}). When $d=n-3$, the ordered partition gives the $(n-2)!$ permutations of $\{2,3,\ldots,n-1\}$; when $d=1$, it gives has $2^{n-2}-2$ terms. Generally, the unordered partition of $n-2$ labels $\{2,3,\ldots,n-1\}$ into $d+1$ nonempty subsets is given by the second kind of Stirling number $S(n-2,d+1)$ \cite{wiki:Stirling_numbers_of_the_second_kind}. For example, $S(n-2,n-2)=1$ and $S(n-2,2)=2^{n-3}-1$. After considering the ordering between these sets, there are $(d+1)!S(n-2,d+1)$ ordered partitions with length $d+1$. Since the partitions are related to the boundaries of permutohedron via \eqref{eq: lBDt}, $(d+1)!S(n-2,d+1)$ also counts the co-dimension $d$ boundaries of permutohedron $\mathcal{P}_{n-2}$. Thus, the total number of partitions with length $1,2,\ldots,n-2$ is the Fubini number $\mathcal{F}_{n-2}=\sum_{d=1}^{n-2}d!S(n-2,d)$ \cite{Mezo}, which is also the total number of all co-dimension boundaries of $\mathcal{P}_{n-2}$.

To make the quasi-shuffle algebra a bialgebra, we can also define the coproduct $\delta: V \rightarrow V\otimes V$ as a linear map, which satisfies \cite{Hoffman, Brandhuber:2021bsf, Brandhuber:2022enp}
\begin{align}
    &\delta((\sigma_0))=\mathbb{I}\otimes(\sigma_0)+(\sigma_0)\otimes \mathbb{I}\\\nonumber
    &\delta((\sigma_0,\sigma_1,\ldots,\sigma_s)\star(\rho_0,\rho_1,\ldots,\rho_t))
=\delta((\sigma_0,\sigma_1,\ldots,\sigma_s))\star\delta((\rho_0,\rho_1,\ldots,\rho_t)).
\end{align}
The definition of the coproduct can also be extended to be consistent with the tensor product as $(A\otimes B)\star(C\otimes D)=(A\star C)\otimes(B\star D)$. Additionally, the unit element of coalgebra $\epsilon$ can be defined as 
\begin{equation}
\epsilon(\mathbb{I})=\mathbb{I},\qquad\epsilon(\sigma)=0.
\end{equation}
To illustrate, let's give an example of the coproduct:
\begin{align}
    \delta(2\star3)&=\mathbb{I}\otimes\left[(2,3)+(3,2)-(23)\right]\\\nonumber
    &+(2)\otimes(3)+(3)\otimes(2)\\\nonumber
    &+\left[(2,3)+(3,2)-(23)\right]\otimes\mathbb{I}
\end{align}
From this example, we can see that the coproduct of the specific quasi-shuffle $(2\star3\star\ldots\star n{-}1)$ which is defined as $\hat{K}(2,3,\ldots,n{-}1)$ can be written in terms of the tensor product of lower-point ones. Generally, we have
\begin{equation}
    \delta(\hat{K}(2,3,\ldots,n{-}1))=\sum_{I\subset\{2,3,\ldots,n{-}1\}}\hat{K}(I){\otimes} \hat{K}(\bar{I}),
\end{equation}
where the summation runs over all subsets of $\{2,3,\ldots,n{-}1\}$ which is allowed to be the empty set, and we define $\delta(\emptyset)=\mathbb{I}$.

Notice that each term in the coproduct of the BCJ numerator factorizes into the product of two lower-point numerators in the sense of \eqref{eq: Kfac}, so it motivates us to define a replacement rule $C$ for tensor products as
\begin{equation}
    C(\hat{K}(I){\otimes} \hat{K}(\bar{I}))=K(1,\mathrm{Id}(I),P) \tilde{K}^I(1,\mathrm{Id}(\bar{I}),n),
\end{equation}
where $\sigma$ and $\rho$ are words. Then the factorization property \eqref{eq: Kfac} can be written in the language of coproduct
\begin{equation}
\left.\mathrm{Res}\right|_{D_I{=}0}K( 1,2,\ldots,n)=\left.\mathrm{Res}\right|_{D_I{=}0} C\ \delta (\hat{K}(2,3,\ldots,n{-}1).
\end{equation}

We can also promote the bialgebra to be a Hopf algebra by defining the antipode map $S: V\rightarrow V$ \cite{Hoffman, Brandhuber:2021bsf, Brandhuber:2022enp}, which satisfies $\star(\mathbb{I}\otimes S)\delta((\sigma))=\star (S\otimes\mathbb{I})\delta((\sigma))=\epsilon(\sigma)\mathbb{I}$, where $\star(\sigma_0 \otimes \sigma_1)\equiv \sigma_0 \star \sigma_1$ . To be explicit, it can be defined recursively
\begin{align}
    &S(\mathbb{I}):=\mathbb{I}\\\nonumber
    &S((\sigma_1,\sigma_2,\ldots,\sigma_r)):=-\sum_{i=0}^{r{-}1}S((\sigma_1,\sigma_2,\ldots,\sigma_i))\star S((\sigma_{i{+}1},\sigma_{i{+}2},\ldots,\sigma_r)).
\end{align}
When acting on the $\hat{K}(2,3,\ldots,n{-}1)$, it trivially gives
\begin{equation}
    S\hat{K}(2,3,\ldots,n{-}1)=(-1)^{n}\hat{K}(2,3,\ldots,n{-}1).
\end{equation}
Thus the antipode only changes the numerator $\hat{K}(2,3,\ldots,n-1)$ by an overall sign. So there is no useful interpretation of the antipode map.

\section{Explicit BCJ numerators for five-point}\label{sec: N5}
For completeness, we provide another explicit example for the five-point BCJ numerator of YMS amplitude, which corresponds to all boundaries of ${\cal P}_3$
\begin{equation}\label{eq: N5}
\begin{aligned}
    &K(1,2,3,4,5) \\
    =&\frac{1}{D_{234}} \left( p_1\cdot F_{234}\cdot q_{234} +\frac{p_1\cdot F_{34}\cdot q_{234}  p_1\cdot F_2\cdot q_2  }{D_2}+\frac{ p_1\cdot F_{24}\cdot q_{234} p_{12}\cdot F_3\cdot q_3 }{D_3}+\frac{ p_1\cdot F_{23}\cdot q_{234} p_{123}\cdot F_4\cdot q_4 }{D_4}\right. \\
    &+\frac{p_1\cdot F_4\cdot q_{234} p_1\cdot F_{23}\cdot q_{23}  }{D_{23}}+\frac{ p_1\cdot F_3\cdot q_{234} p_1\cdot F_{24}\cdot q_{24} }{D_{24}}+\frac{ p_1\cdot F_2\cdot q_{234} p_{12}\cdot F_{34}\cdot q_{34} }{D_{34}}\\
    &+\frac{ p_1\cdot F_4\cdot q_{234} p_1\cdot F_3\cdot q_{23} p_1\cdot F_2\cdot q_2  }{D_{23}  D_2  }+\frac{   p_1\cdot F_4\cdot q_{234} p_1\cdot F_2\cdot q_{23} p_{12}\cdot F_3\cdot q_3 }{D_{23}  D_3  }\\
    &+\frac{ p_1\cdot F_3\cdot q_{234}  p_{13}\cdot F_4\cdot q_{24}   p_1\cdot F_2\cdot q_2  }{D_{24}  D_2  }+\frac{  p_1\cdot F_3\cdot q_{234}  p_1\cdot F_2\cdot q_{24} p_{123}\cdot F_4\cdot q_4 }{D_{24} D_4  } \\
    &\left.+\frac{p_1\cdot F_2\cdot q_{234} p_{12}\cdot F_4\cdot q_{34}   p_{12}\cdot F_3\cdot q_3  }{ D_{34} D_3  }  
    +\frac{p_1\cdot F_2\cdot q_{234} p_{12}\cdot F_3\cdot q_{34}  p_{123}\cdot F_4\cdot q_4}{D_{34} D_4 }\right)
\end{aligned}
\end{equation}
It contains $13$ terms, or the Fubini number $\mathcal{F}_3$. These $13$ terms can also be realized by quasi-shuffle product $2\star 3\star 4$ evaluated as~\eqref{eq: part} acted by the linear map~\eqref{eq:linmap}: in the same order as the above equation, the partitions are $(\{234\})$, $(\{34\}, \{2\})$, $(\{24\},\{3\})$, $(\{23\},\{4\})$, $(\{4\},\{23\})$, $(\{3\},\{24\})$, $(\{2\},\{34\})$, $(\{4\},\{3\},\{2\})$, $(\{4\},\{2\},\{3\})$, $(\{3\},\{4\},\{2\})$, $(\{3\},\{2\},\{4\})$, $(\{2\},\{4\},\{3\})$, $(\{2\},\{3\},\{4\})$.

The five-point YM numerator $K^{\text{YM}}(1,2,3,4,5)$ is almost the same as \eqref{eq: N5} except that with our special choice, in each term the factor $p_1\cdot F_{\tau_0}\cdot q_{234}$ is changed to $\varepsilon_5\cdot F_{\tau_0}\cdot\varepsilon_5$ and the overall pole $D_{234}$ is changed to $\varepsilon_5\cdot p_{234}=-\varepsilon_5\cdot p_1$.

\section{Details of the effective numerators in heavy limit and double copy}\label{sec:heavy}
\paragraph{Effective BCJ numerators for HEFT} Recall in \eqref{eq: Kexpand} we expand $K_\mathrm{H}$ into $K_\mathrm{H}^{(i)}$ by counting the power of $p_1$; therefore we have a good control of the maximal power of the heavy mass $m$. For $n=4,5$, the expansion is trivial since we have one and three terms all contribute to $K_\mathrm{H}^{(2)}(1,2,3,4)$ and $K_\mathrm{H}^{(2)}(1,2,3,4,5)$ respectively as given in sec.\ref{sec:heavy1}. For $n=6$, the following term with explicit factor $m^2$ will contribute to $K^{(2)}_{H}(1,2,\ldots,6)$: 
\begin{equation}
   -\frac{2m^2}{k^2} \frac{v\cdot F_{23} \cdot v\  p_{23}\cdot F_5 \cdot v\ p_{23}\cdot F_4 \cdot v } { v \cdot p_{45}\ v\cdot p_4 }.
\end{equation}
Meanwhile $K^{(3)}_{\rm H}(1,2,\ldots,6)$ is given by
\begin{equation} \label{eq: 6ptKH3}
\begin{aligned}
-\frac{2m^3}{k^2}&\left(\frac{v\cdot F_{25}\cdot v\ v\cdot F_{34}\cdot v}{v\cdot p_{34}}+\frac{v\cdot F_{25}\cdot v\ v\cdot F_{34}\cdot v}{ v\cdot p_{25}}+ \frac{v\cdot F_{24}\cdot v\ v\cdot F_{35}\cdot v}{ v\cdot p_{24}}\right. \\
&\left. +\frac{v\cdot F_{24}\cdot v\ v\cdot F_{35}\cdot v}{ v\cdot p_{35}}+
\frac{v\cdot F_{23}\cdot v\ v\cdot F_{45}\cdot v}{ v\cdot p_{23}}+\frac{v\cdot F_{23}\cdot v\ v\cdot F_{45}\cdot v}{ v\cdot p_{45}}\right).
\end{aligned}
\end{equation}
Importantly, the actual power of $m$ for $K_\mathrm{H}^{(i)}$ is $2$ for $i=2$ and $i-1$ for $2<i\leq \lfloor n/2 \rfloor$ since for the latter cases the expressions are proportional to an addition $v\cdot k \propto m^{-1}$ after collecting terms carefully. For instance, \eqref{eq: 6ptKH3} can be rewritten as
\begin{equation}
    -\frac{2m^3}{k^2} v\cdot k \left(\frac{v\cdot F_{25}\cdot v\ v\cdot F_{34}\cdot v}{v\cdot p_{34} v\cdot p_{25}} +\frac{v\cdot F_{24}\cdot v\ v\cdot F_{35}\cdot v}{ v\cdot p_{24} v\cdot p_{35}}+\frac{v\cdot F_{23}\cdot v\ v\cdot F_{45}\cdot v}{ v\cdot p_{23}v\cdot p_{45}}  \right)
\end{equation}


On the other hand, we also expand $\phi^3$ amplitudes according to the order of $m^{-1}$,
\begin{equation}
    A^{\phi^3}(1,\beta,n)= \sum_{j=0}^{n-3} A^{\phi^3,(j)}(1,\beta,n),
\end{equation}
where $ A^{\phi^3,(j)}(1,\beta,n)$ denoted terms with the order $\mathcal{O}(m^{-j})$. For instance, at $n=6$ and $j=2,3$ we have
\begin{equation} 
\begin{aligned}
    A^{\phi^3,(2)}(1,2,\ldots,6)=\frac{1}{s_{12} s_{34} s_{1234}}+\frac{1}{s_{23} s_{123} s_{1234}}+\frac{1}{s_{12} s_{45} s_{123}}, \
    A^{\phi^3,(3)}(1,2,\ldots,6)=\frac{1}{s_{12} s_{123} s_{1234}}.
\end{aligned}
\end{equation}
Therefore the amplitudes are expressed by
\begin{equation} \label{eq:Hampexpand}
\begin{aligned}
    A^\mathrm{H}(1,2,\ldots,n)=& \sum_{\beta \in S_{n-2}}A^{\phi^3}(1,\beta,n) K_{\mathrm{H}}(1,\beta,n) \\
    =&\sum^{n-3}_{j=0} \sum_{\Theta^j} \sum_{i=2}^{\lfloor n/2 \rfloor } \frac{K_{\mathrm{H}}^{(i)}(1,\Theta^j,n)}{d_{\Theta^{j}}},
\end{aligned}
\end{equation}
Here, each term in the summation is at order $\mathcal{O}(m^{i-j})$ for $i=2$ and $\mathcal{O}(m^{i-1-j})$ otherwise. We define a nested commutator of depth $r$ of an ordered set to be $r$ mutually compatible commutators acting on the ordered set; and we use $\Theta^j$ for $j=0,1,\ldots,n-3$ to represent nested commutator of depth $n-3,n-4,\ldots,0$ of the ordered set $(2,3,\ldots,n-1)$. For examples, for $n=5$ with $j=0,1,2$ we sum over
$\Theta^0=([[2,3],4]), ([2,[3,4]])$ $\Theta^1=([2,3],4),( 2,[3,4])$ and $\Theta^2=(2,3,4)$.
Moreover, $d_{\Theta^{j}}$ is the propagator denominator corresponding to the cubic graph associated with $\Theta^j$:
\begin{equation*}
    \begin{aligned}\includegraphics[width=0.2\linewidth]{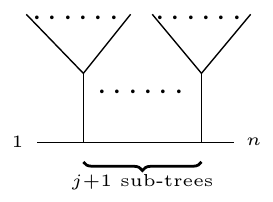}\end{aligned} 
 \ \leftrightarrow \ d_{\Theta^j}
\end{equation*}
For $n=5$, $d_{\Theta^{j}}$'s involved in the summation are given by
\begin{equation*}
    \begin{aligned}\includegraphics[width=0.2\linewidth]{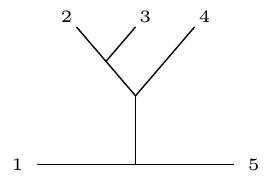}\end{aligned} \leftrightarrow d_{[[2,3],4]}=s_{23}s_{234},     \begin{aligned}\includegraphics[width=0.2\linewidth]{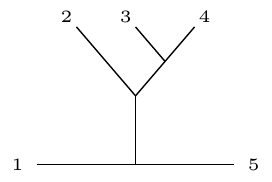}\end{aligned} \leftrightarrow d_{[2,[3,4]]}=s_{34}s_{234},
\end{equation*}

\begin{equation*}
    \begin{aligned}\includegraphics[width=0.2\linewidth]{figs/theta511.pdf}\end{aligned} \leftrightarrow d_{[2,3],4}=s_{123}s_{23}, 
    \begin{aligned}\includegraphics[width=0.2\linewidth]{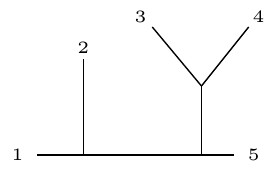}\end{aligned} \leftrightarrow d_{2,[3,4]}=s_{12}s_{34}, 
\end{equation*}

\begin{equation*}
    \begin{aligned}\includegraphics[width=0.2\linewidth]{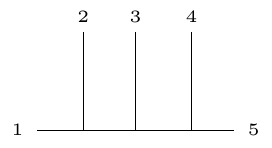}\end{aligned} \leftrightarrow d_{2,3,4}=s_{12}s_{123}.
\end{equation*}



Notice that the correct order of the amplitude $A^\mathrm{H}(1,2,\ldots,n)$ is $\mathcal{O}(m^1)$, therefore physically one expects the contribution to higher power of $m$ vanishes. We have checked this fact up to $n=10$. Moreover, we have observed that the only contribution to the leading order of amplitudes is $i=2$ and $j=1$, which means 
\begin{equation} \label{eq: KHcomvanish}
    K_\mathrm{H}^{(2)}(\Theta^0)=0\text{ and } K_{\mathrm{H}}^{(i)}(1,\Theta^{i-2},n)=0 \text{ for } 2<i\leq \lfloor n/2 \rfloor.
\end{equation}
Note that 
\begin{equation} \label{eq:vanishtovanish}
    K_{\mathrm{H}}^{(i)}(1,\Theta^{j},n)=0 \Rightarrow  K_{\mathrm{H}}^{(i)}(1,\Theta^{a},n)=0 \text{ for $a <j$},
\end{equation}
since $K_{\mathrm{H}}^{(i)}(1,\Theta^{a},n)$ is nothing but the linear combination of those with $j>a$. Therefore, equations ~\eqref{eq: KHcomvanish} already insure the vanishing of $\mathcal{O}(m^h)$ with $h>1$.
In addition, the general conjecture which is even stronger than \eqref{eq: KHcomvanish} and confirmed up to 10-point reads
\begin{equation} \label{eq:KHconjecture}
    K_{\mathrm{H}}^{(i)}(1,\Theta^{2(i-2)},n)=0 \text{ for } 2\leq i\leq \lfloor n/2 \rfloor.
\end{equation}
We now give a proof of the simplest case of conjecture~\eqref{eq:KHconjecture}, say $i=n/2$ for even $n$, for which we have
\begin{equation}
    K_{\mathrm{H}}^{(n/2)}(1,2,\ldots,n)= \sum_{\tau \in \mathrm{part}^{ (n/2)}, |\tau_k|=2 } \prod_{k=0}^{n/2-1}  \frac{p_1 \cdot F_{\tau_k} \cdot v} {D_{I_k}},
\end{equation}
where we sum over the ordered partition of $(2,\ldots,n-1)$ into $n/2$ sets with each set contains two elements. The above expression is invariant under the permutation on any two labels, therefore
\begin{equation}
    K_{\mathrm{H}}^{(n/2)}(1,\Theta^{n-4},n)=0,
\end{equation}
which completes the proof.

In summary, the effective BCJ numerator contributing to the leading order of the amplitude $A^\mathrm{H}(1,2,\ldots,n)$ is $K^{(2)}_{\mathrm{H}}$ where the power of $m$ is only contributed from the overall pole $v\cdot k=-\frac{k^2}{2m}$ and the factor $m v\cdot F_{\tau_0}\cdot v$ in numerator~\eqref{eq: KBD}. 
This fact is equivalent to the boundary \eqref{eq: lBD} does not contain particle $2$ except for the $I_{0}$. The proof is straightforward. Whenever there is a $I_{k}(k\neq d)$ which consists of $2$ and a $I_{k{+}1}$ does not consist of $2$, then $\Delta(I_{k},I_{k+1})=\left.\bar{I_k}\right|_{<2}=\emptyset$, which means that this boundary at least contributes to the superficial order of $\mathcal{O}(m^3)$, unless $\abs{\tau_k}=1$ for which this term vanishes. Similarly, for the remaining $I_d\supset\{2\}$ case, $\Delta(I_d,\emptyset)$ also becomes empty-set, and this boundary will either contribute to at least superficial order $\mathcal{O}(m^3)$ or vanishes.

Therefore, the remained pole structure contains two parts of contribution. The pole structure of the first part is related to $\mathcal{P}_{\{34\ldots n{-}1\}}$ with an additional overall pole $D_{23\ldots n{-}1}$, which implies that there are $\mathcal{F}_{n{-}3}$ terms. The other part is also corresponded to $\mathcal{P}_{\{34\ldots n{-}1\}}$, but does not have the pole $D_{34\ldots n{-}1}$. So naively there are $2\mathcal{F}_{n{-}3}$ terms contributing to order $\mathcal{O}(m^1)$. However, for the first part which contains pole $D_{34\ldots n{-}1}$, the numerator of each term must include a factor $m v\cdot F_2\cdot v$ hence vanishes. Consequently, we obtain the expression of the effective BCJ numerator~\eqref{eq: BDexph} in the heavy limit and the leading order of amplitude is given by \eqref{eq: Hamp}.

\paragraph{Double copy} It is quite interesting that the vanishing properties~\eqref{eq:KHconjecture} are precisely sufficient to ensure the leading order of amplitude for heavy scalars coupled to gravitons also only receives the contribution from $K_{\mathrm{H}}^{(2)}$. Consider the double copy of~\eqref{eq:Hampexpand}
\begin{equation} 
\begin{aligned}
    M^{\rm H}_n=& \sum_{\alpha,\beta \in S_{n-2}}K_{\mathrm{H}}(1,\alpha,n) m(1,\alpha,n|1,\beta,n) K_{\mathrm{H}}(1,\beta,n) \\
    =&\sum^{n-3}_{j=0} \sum_{\Theta^{\prime j}} \sum_{i,i^\prime =2}^{\lfloor n/2 \rfloor } \frac{K_{\mathrm{H}}^{(i)}(1,\Theta^{\prime j},n)K_{\mathrm{H}}^{(i^\prime)}(1,\Theta^{\prime j},n)}{d_{\Theta^{\prime j}}},
\end{aligned}
\end{equation}
where $\Theta^{\prime j}$ for $j=0,1,\ldots,n-3$ is the nested commutators of depth $n-3,n-2,\ldots,0$ of the unordered set $\{2,3,\ldots,n-1\}$ and $d_{\Theta^{\prime j}}$ is the corresponding propagator denominator. Note that on the support of~\eqref{eq:vanishtovanish}, conjecture~\eqref{eq:KHconjecture} implies
\begin{align}
    & K_{\mathrm{H}}^{(i)}(1,\Theta^{\prime j},n)K_{\mathrm{H}}^{(i^\prime)}(1,\Theta^{\prime j},n)=0 \text{ for } i+i^\prime-j-2=2 \text{ with } i,i^\prime \neq 2 \\
    & K_{\mathrm{H}}^{(2)}(1,\Theta^{\prime j},n)K_{\mathrm{H}}^{(i^\prime)}(1,\Theta^{\prime j},n)=0 \text{ for } i^\prime-j+1=2 \text{ with } i^\prime \neq 2, 
\end{align}
which would have contributed at $\mathcal{O}(m^2)$. Therefore, the amplitude is simplified into
\begin{equation} \label{eq:HampGR}
    M^{\rm H}_n= \sum_{\Theta^{\prime 1} } \frac{K_\mathrm{H}^{(2)}(1,\Theta^{\prime 1} ,n)^2}{d_{\Theta^{\prime 1}}} + \sum_{\Theta^{\prime 2} } \frac{K_\mathrm{H}^{(2)}(1,\Theta^{\prime 2} ,n)^2}{d_{\Theta^{\prime 2}}},
\end{equation}
Importantly, the above two contributions are at the same order $\mathcal{O}(m^2)$, where for the first part, {\it i.e.} the contribution from summing over $\Theta^{\prime 1}$, one needs to collect terms in pair on the support of~\eqref{eq:KHconjecture} to organize the result in explicit $\mathcal{O}(m^2)$. For $n=4$, there would be no contribution from the second part and the amplitude reads
\begin{equation}
    M^{\rm H}_4= \frac{K_\mathrm{H}^{(2)}(1,2,3,4)^2}{s_{12}} +\frac{K_\mathrm{H}^{(2)}(1,3,2,4)^2}{s_{13}}=  \frac{2m v\cdot k\ K_\mathrm{H}^{(2)}(1,2,3,4)^2}{ s_{12}s_{13}},
\end{equation}
where we have used $K_\mathrm{H}^{(2)}(1,2,3,4)=K_\mathrm{H}^{(2)}(1,3,2,4)$. A more nontrivial case is for $n=5$, the result is given by
\begin{equation}\label{eq:HampGR5}
\begin{aligned} 
    &\frac{1}{s_{34}} \left( \frac{K_\mathrm{H}^{(2)}(2,[3,4])^2}{s_{12}} +\frac{K_\mathrm{H}^{(2)}([3,4],2)^2}{s_{134}} \right) + \frac{1}{s_{24}} \left( \frac{K_\mathrm{H}^{(2)}(3,[2,4])^2}{s_{13}} +\frac{K_\mathrm{H}^{(2)}([2,4],3)^2}{s_{124}} \right)+ \frac{1}{s_{23}} \left( \frac{K_\mathrm{H}^{(2)}(4,[2,3])^2}{s_{14}} +\frac{K_\mathrm{H}^{(2)}([2,3],4)^2}{s_{123}} \right) \\
    &+\frac{K_\mathrm{H}^{(2)}(2,3,4)^2}{s_{12}s_{123}} +\frac{K_\mathrm{H}^{(2)}(3,2,4)^2}{s_{13}s_{123}}
    +\frac{K_\mathrm{H}^{(2)}(2,4,3)^2}{s_{12}s_{124}}
    +\frac{K_\mathrm{H}^{(2)}(4,2,3)^2}{s_{14}s_{124}}
    +\frac{K_\mathrm{H}^{(2)}(3,4,2)^2}{s_{13}s_{134}}
    +\frac{K_\mathrm{H}^{(2)}(4,3,2)^2}{s_{14}s_{134}},
\end{aligned}
\end{equation}
where we have omitted the scalar labels $1$ and $5$ in $K_\mathrm{H}^{(2)}$. Note that for two terms in each pair in the first line, the numerators are equal which is the part of the conjecture~\eqref{eq:KHconjecture} with $i=2,j=0$. Therefore, by collecting terms in pairs the first line gives rise to an additional $v\cdot k$ and we have the amplitude consistently at $\mathcal{O}(m^2)$. This argument works for general $n$ with each pair corresponding to cubic graphs with two identical sub-trees placed reversely on the scalar line.

\begin{figure}[H]
\centering
    \centering
    \subfloat{ 
    \begin{minipage}[c]{0.3\linewidth}
    \centering
    \includegraphics[scale=1.3]{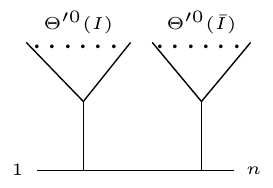}
    \end{minipage}
    }
    \subfloat{ 
    \begin{minipage}[c]{0.3\linewidth} 
    \centering
    \includegraphics[scale=1.3]{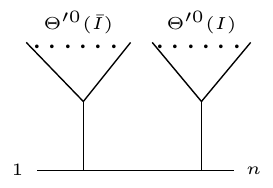}
    \end{minipage}
    }
\caption{Cubic graphs with two identical sub-trees placed reversely on the scalar line.}
\label{fig:2cubicg}
\end{figure}

Moreover, the amplitude~\eqref{eq:HampGR} can be simplified further, {\it e.g.} for~\eqref{eq:HampGR5}, the first pair in the first line reads
\begin{equation}
    \frac{1}{s_{34}} \frac{(s_{12}+s_{134})K_\mathrm{H}^{(2)}(2,[3,4])^2}{s_{12}s_{134}}=\frac{1}{s_{34}} \frac{(2mv\cdot k + s_{34})K_\mathrm{H}^{(2)}(2,[3,4])^2}{s_{12}s_{134}}.
\end{equation}
Similar operations can be applied to the remaining two pairs, and one can check that the amplitude is then given by
\begin{equation}
    M^{\rm H}_5=2mv\cdot k \left( \frac{K_\mathrm{H}^{(2)}(2,[3,4])^2}{s_{12}s_{134}s_{34}} + \frac{K_\mathrm{H}^{(2)}(3,[2,4])^2}{s_{13}s_{124}s_{24}} + \frac{K_\mathrm{H}^{(2)}([2,3],4)^2}{s_{14}s_{123}s_{23}} \right),
\end{equation}
where the second line of~\eqref{eq:HampGR5} has been canceled. The analogous cancellation has also been observed at higher points, which leads to the conjecture of a more compact version of~\eqref{eq:HampGR}
\begin{equation} 
    M^{\rm H}_n= 2mv\cdot k \sum_{ g } \frac{K_\mathrm{H}^{(2)}(1,\Theta^{\prime 0}(I),\Theta^{\prime 0}(\bar{I}),n)^2}{s_{1I}s_{1\bar{I}}d_{\Theta^{\prime 0}(I)}d_{\Theta^{\prime 0}(\bar{I})}},
\end{equation}
where we sum over the cubic graphs $g$ corresponding to half of the possible $\Theta^{\prime 1}$'s, {\it i.e.} each graph is either of two graphs with two identical sub-trees (contain legs $I$ and $\bar{I}$) where the sub-trees are placed reversely on the scalar line (see figure~\ref{fig:2cubicg}). We also define $\Theta^{\prime 0}(I)$ to be the nested commutator of depth $|I|-1$ of the unordered set $I$ and $d_{\Theta^{\prime 0}(I)}$ to be the corresponding propagator denominator; for each term in the summation, $\Theta^{\prime 0}(I)$ and $\Theta^{\prime 0}(\bar{I})$ are given by the sub-trees of the graph $g$. The above expression contains $(2n-7)!!$ terms, which is equal to the number of terms in an $(n-1)$-point amplitude of gravitons scattering(without heavy scalars).
    
\bibliographystyle{apsrev4-1}
\bibliography{bib}

\end{document}